\documentclass{aastex6}

\slugcomment{Accepted for publication in ApJ}

\begin{document}


\title{Doubly imaged quasar SDSS J1515+1511: time delay and lensing galaxies}
\shorttitle{SDSS J1515+1511: time delay and lensing galaxies}



\author{Vyacheslav N. Shalyapin\altaffilmark{1,2} and Luis J. Goicoechea\altaffilmark{2}}

\altaffiltext{1}{Institute for Radiophysics and Electronics, National Academy of Sciences 
of Ukraine, 12 Proskura St., 61085 Kharkov, Ukraine; vshal@ukr.net} 
\altaffiltext{2}{Departamento de F\'\i sica Moderna, Universidad de Cantabria, Avda. de 
Los Castros s/n, 39005 Santander, Spain; goicol@unican.es}

\begin{abstract}
We analyse new optical observations of the gravitational lens system SDSS J1515+1511. These include 
a 2.6--year photometric monitoring with the Liverpool Telescope (LT) in the $r$ band, as well as a 
spectroscopic follow--up with the LT and the Gran Telescopio Canarias (GTC). Our $r$--band LT light 
curves cover a quiescent microlensing period of the doubly imaged quasar at $z_{\rm{s}}$ = 2.049, 
which permits us to robustly estimate the time delay between the two images A and B: 211 $\pm$ 5 
days (1$\sigma$ confidence interval; A is leading). Unfortunately, the main lensing galaxy (G1) is 
so faint and close to the bright quasar that it is not feasible to accurately extract its spectrum 
through the GTC data. However, assuming the putative redshift $z_{\rm{G1}}$ = 0.742, the GTC and LT 
spectra of the distant quasar are used to discuss the macrolens magnification, and the extinction 
and microlensing effects in G1. The new constraints on the time delay and macrolens magnification 
ratio essentially do not change previous findings on the mass scale of G1 and external shear, while 
the redshift of the lensing mass is found to be consistent with the assumed value of $z_{\rm{G1}}$. 
This is a clear evidence that G1 is indeed located at $z_{\rm{G1}}$ = 0.742. From the GTC data we 
also obtain the redshift of two additional objects (the secondary galaxy G2 and a new absorption 
system) and discuss their possible role in the lens scenario. 
\end{abstract}


\section{Introduction} \label{sec:intro}

Optical light curves and spectra of gravitationally lensed quasars show different phases of 
microlensing activity induced by stars in lensing galaxies, and each of these phases is relevant 
for certain astrophysical studies. Periods of high microlensing activity were reported for several 
systems, including prominent flux variations \citep[e.g.,][]{2006AcA....56..293U,
2013ApJ...774...69H}, significant spectral distortions \citep[e.g.,][]{2004ApJ...610..679R,
2014ApJ...797...61R}, or both \citep[e.g.,][]{goico2016}. Such high activity is related to stellar 
mass microlenses strongly affecting the accretion disk and the innermost part of the broad line 
region (BLR), while the narrow line region (NLR) remains unaffected 
\citep[e.g.,][]{1990A&A...237...42S,2002ApJ...576..640A}. Hence, a strong microlensing episode in a 
lensed quasar can be used to probe the structure of its accretion disk and BLR, as well as the 
composition of the main lensing galaxy \citep[e.g.,][and references therein]{2011ApJ...738...96M}. 
Unfortunately, when an important extrinsic (microlensing) variability is present in the light curve 
of an image of a lensed quasar, it is difficult to accurately measure the time delay between that 
image and any other. Only in some cases, time delays were measured to $\sim$ 5\% precision after 
monitoring during 5$-$10 year periods and using sophisticated techniques of analysis 
\citep[e.g.,][]{2003ApJ...594..101O,2008ApJ...676...80M,2013ApJ...774...69H}.

Time delays can be used to constrain the Hubble constant (and other cosmological parameters) and 
lensing mass distributions \citep[e.g.,][]{1964MNRAS.128..307R,1966MNRAS.132..101R,
2006glsw.conf.....M}. Therefore, lensed quasars with two or more images in quiescent phases of 
microlensing activity are ideal targets to determine delays and study the Universe on different 
scales. Knowing the time delay between two quiescent images of the same quasar, if we detect a 
sharp variation in the optical flux of the leading image, a multiwavelength campaign can be planned 
to follow the variability of the trailing one. Multiwavelength intrinsic variations in this 
trailing image are valuable tools to carry out reverberation mapping studies 
\citep[e.g.,][]{2012ApJ...744...47G,2015ApJ...813...67D}, and thus check the quasar structure 
derived from active microlensing periods. Time delays from quiescent images and images showing 
smooth microlensing variations are often measured to $\leq$ 3\% precision \citep[e.g.,][]{
2013A&A...553A.121E,2013A&A...557A..44R}, and sometimes to precision levels as low as a few tenths 
of a percent \citep[e.g.,][]{1997ApJ...482...75K,2008A&A...492..401S}.

The lensed quasar \object{SDSS J1515+1511} consists of two optically bright images (A and B) 
separated by $\sim 2\arcsec$ \citep{2014AJ....147..153I}. Although Inada et al. reported a source 
redshift $z_{\rm{s}}$ = 2.054, this relies on the SDSS spectrum of the A image taken in 2007, which 
has all its emission features at $z_{\rm{em}}$ = 2.049. It is also worth to note that the SDSS/BOSS 
spectrum of A in 2012, as well as the new spectra of A and B in 2015 (see Section 
\ref{subsubsec:gtcspec}), contain emission lines at a redshift of 2.049. We thus adopt $z_{\rm{s}}$ 
= $z_{\rm{em}}$ = 2.049 throughout this paper. Inada et al. also presented spectra of A and B that
were obtained through a 900 s exposure with the DOLORES spectrograph on the 3.6 m Telescopio 
Nazionale Galileo (TNG). They found strong Mg\,{\sc ii} absorption at $\sim$4900 \AA\ in the 
spectrum of B (the image closer to the main lensing galaxy G1). Such an absorption feature 
corresponds to intervening gas at a redshift of 0.742, suggesting that $z_{\rm{G1}}$ = 0.742. New 
Subaru Telescope adaptive--optics observations in the $K'$--band improved the relative astrometry 
of the lens system and the morphological parameters of G1 \citep{2016MNRAS.458....2R}. Rusu et al. 
also detected a secondary galaxy G2 southwest of the ABG1 system and predicted the time delay 
$\Delta t_{\rm{AB}}$ for three different lens models. To perform this model fitting, the macrolens 
magnification ratio $\Delta m_{\rm{AB}}$ was constrained within an interval based on the $K'$--band 
magnitude difference $B - A$. The expected value of $\Delta t_{\rm{AB}}$ ranged from $\sim$ 145 to 
$\sim$ 216 days, with the longest delay for the most realistic model: singular isothermal ellipsoid 
plus external shear (SIE+$\gamma$). 

Here we focus on observations of \object{SDSS J1515+1511} during a quiescent phase of microlensing 
activity over the 2014$-$2016 period. The outline of the paper is the following: Section 
\ref{sec:data} presents a 2.6--year photometric monitoring of both quasar images, and spectroscopic 
follow--up observations of the lens system. These last observations include new long--slit spectra
that are 50 times deeper than the previous ones with the TNG (see above). Section \ref{sec:delay} 
is devoted to accurately measure $\Delta t_{\rm{AB}}$, while we estimate $\Delta m_{\rm{AB}}$ in 
Section \ref{sec:extmagmic}. In Section \ref{sec:extmagmic}, we also discuss the dust extinction 
and microlensing magnification in G1. In Section \ref{sec:lensmod}, we update the SIE+$\gamma$ lens
model, confirm the tentative redshift of G1, and discuss the role of G2 (and other intervening 
object) in the lensing phenomenon. Our main results and conclusions appear in Section 
\ref{sec:end}.  

\section{Observations and data reduction} \label{sec:data}

\subsection{Photometric monitoring} \label{subsec:livmon}

We conducted a photometric monitoring campaign of \object{SDSS J1515+1511} from early February of 
2014 to mid September of 2016, i.e., for 2.6 years with an average sampling rate of about two 
nights every week. All optical observations were performed with the 2.0 m fully robotic Liverpool 
Telescope (LT) at the Roque de los Muchachos Observatory, Canary Islands (Spain), using the IO:O 
CCD camera (pixel scale of $0\farcs 30$). Each observing night, 2 $\times$ 300 s exposures were 
taken in the $r$ Sloan passband, and the corresponding frames were subsequently passed through a 
pre--processing pipeline. The LT data reduction pipeline included bias subtraction, overscan 
trimming, and flat fielding. In addition, we cleaned cosmic rays and interpolated over bad pixels 
using the bad pixel mask\footnote{The pre--processed frames will be soon publicly available on the 
GLENDAMA archive at \url{http://grupos.unican.es/glendama/database/} \citep{2015arXiv150504317G}.}. 

\begin{figure}[ht!]
\hspace{-15.00mm}
	\begin{minipage}[h]{0.3\linewidth}
	\centering
      \includegraphics[width=2.0\textwidth]{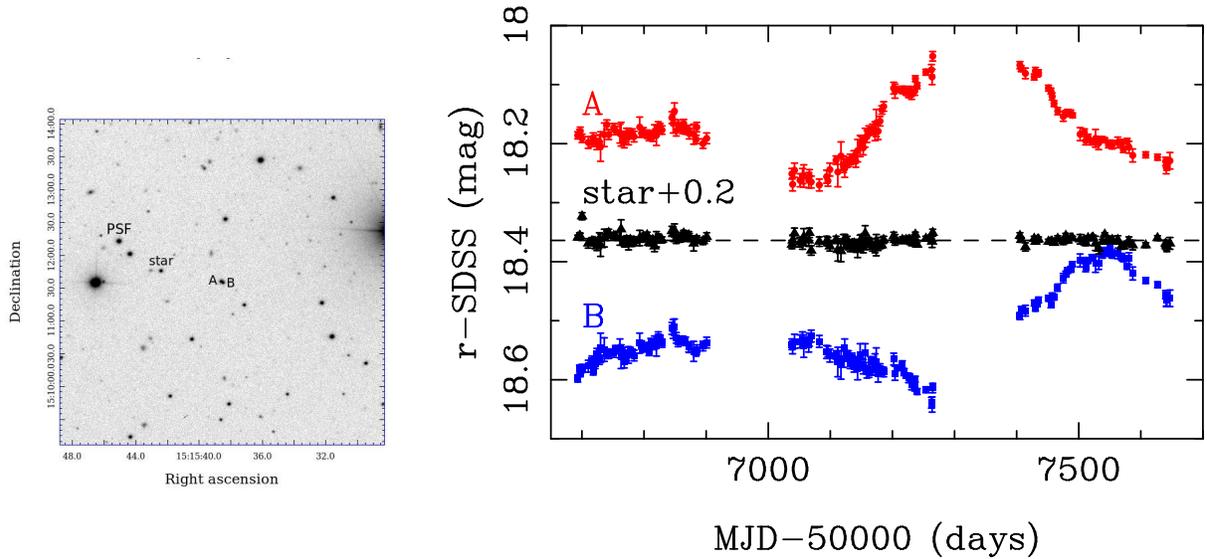}
      \end{minipage}
\hspace{15.00mm}
	\begin{minipage}[h]{0.7\linewidth}
      \centering
      \includegraphics[angle=-90,width=0.8\textwidth]{LCs.eps}
    	\end{minipage}
\caption{Left: First LT frame of SDSS J1515+1511 on 2014 February 4. This 300 s exposure in the $r$ 
band was taken under typical seeing conditions (FWHM = $1\farcs 3$), just a few weeks after the 
announcement of the discovery by N. Inada and coworkers. The field of view covers $5\arcmin \times 
5\arcmin$, and includes the blended quasar images (A and B), the PSF star, and a field star that 
lies about midway between the lensed quasar and the PSF star. Right: LT $r$--band light curves of 
A, B and the field star. The stellar curve is shifted by +0.2 mag to facilitate comparison.}
\label{fig:f1} 
\end{figure}

After the basic reduction of frames, we assumed that the point--spread function (PSF) for each 
exposure is defined by the surface brightness distribution of the unsaturated star with $r$ = 
16.770 mag in the vicinity of the lens system (see the left panel of Figure \ref{fig:f1}). We also 
adopted this PSF star as the reference object for differential photometry. The relative fluxes and 
magnitudes of both quasar images were then derived through PSF fitting. Our crowded--field 
photometry pipeline relied on IRAF\footnote{IRAF is distributed by the National Optical Astronomy 
Observatory, which is operated by the Association of Universities for Research in Astronomy (AURA) 
under cooperative agreement with the National Science Foundation. This software is available at 
\url{http://iraf.noao.edu/}.} tasks and the IMFITFITS software \citep{1998AJ....115.1377M}, while 
the photometric model consisted of two PSFs (A and B), an exponential profile convolved with the 
PSF (G1 galaxy located between A and B), and a constant background. \citet{2014AJ....147..153I} 
reported that the S\'ersic index of G1 is consistent with $n$ = 1 (exponential profile), instead of 
$n$ = 4 (de Vaucouleurs profile). Moreover, \citet{2016MNRAS.458....2R} found that G1 is an 
edge--on disk--like galaxy. The ellipticity, orientation, and effective radius of G1, as well as 
the relative positions of B and G1 (with respect to A), were taken from 
\citet{2016MNRAS.458....2R}. 

The galaxy brightness was estimated from the best frames in terms of signal--to--noise ratio (SNR) 
and full--width at half--maximum (FWHM) of the seeing disc. These frames led to a galaxy--to--PSF 
star ratio $G1/PSF$ = 0.005, or equivalently, $r$(G1) = 22.5 mag. We then applied our pipeline to 
the 310 usable frames, incorporating this $G1/PSF$ value as an additional constraint. As G1 is very 
faint in the $r$ band, and the magnitude difference between G1 and the faintest quasar image is 
$\sim$4 ($B/G1 \sim$ 40), we note that the adopted galaxy model does not play a critical role in 
extracting quasar fluxes. We identified 19 frames producing outliers in the quasar light curves, so 
these were removed from the final database. The remaining 291 frames (average FWHM of $\sim 1\farcs 
4$) allowed us to obtain $r$--SDSS magnitudes of A and B at 150 nights (epochs). We also calculated 
the magnitudes of a field star at these epochs. The field star has a brightness similar to that of 
A, and is located between the lens system and the PSF star (see the left panel of Figure 
\ref{fig:f1}).

\begin{deluxetable}{ccccccc}
\tablecaption{LT $r$--band light curves of SDSS J1515+1511AB.\label{tab:t1}}
\tablenum{1}
\tablewidth{0pt}
\tablehead{
\colhead{MJD--50000} & 
\colhead{A\tablenotemark{a}} & \colhead{$\sigma_{\rm{A}}$\tablenotemark{a}} & 
\colhead{B\tablenotemark{a}} & \colhead{$\sigma_{\rm{B}}$\tablenotemark{a}} & 
\colhead{S\tablenotemark{ab}} & \colhead{$\sigma_{\rm{S}}$\tablenotemark{ab}}
}
\startdata
6693.260 & 18.186 & 0.006 & 18.598 & 0.006 & 18.159 & 0.004 \\
6696.282 & 18.183 & 0.007 & 18.582 & 0.007 & 18.156 & 0.005 \\
6697.263 & 18.179 & 0.007 & 18.588 & 0.007 & 18.157 & 0.005 \\
6700.232 & 18.191 & 0.007 & 18.581 & 0.007 & 18.123 & 0.006 \\
6710.173 & 18.198 & 0.010 & 18.577 & 0.010 & 18.169 & 0.008 \\
\enddata
\tablenotetext{a}{$r$--SDSS magnitude.}
\tablenotetext{b}{We use S to denote the field (control) star.}
\tablecomments{Table 1 is published in its entirety in the machine-readable format.
A portion is shown here for guidance regarding its form and content.}
\end{deluxetable}

To estimate typical photometric errors in the light curves of A, B, and the star, we first 
determined the standard deviations between magnitudes having time separations $<$ 2.5 days, and 
then such deviations were divided by the square root of 2 \citep[e.g.,][]{2010ApJ...708..995G}. 
This procedure led to a typical uncertainty of 0.0085 mag in the brightness of both quasar images, 
which is slightly larger than the typical error in the star (0.0065 mag). We also considered that 
$\sigma_r \propto$ 1/SNR \citep[e.g.,][]{howell06} to infer nightly errors in the light curves. 
Table \ref{tab:t1} includes the final $r$--SDSS magnitudes and errors for A, B and the star. 
These brightness records are also shown in the right panel of Figure \ref{fig:f1}. Apart from the 
two unavoidable seasonal gaps, the quasar variability is accurately traced.

\subsection{Spectroscopic follow--up}

\subsubsection{Deep long--slit spectra} \label{subsubsec:gtcspec}

Deep long--slit spectra of \object{SDSS J1515+1511} were obtained using the R500B and R500R grisms 
in the OSIRIS spectrograph on the 10.4 m Gran Telescopio Canarias (GTC)\footnote{The OSIRIS User 
Manual (v3.1) by A. Cabrera-Lavers is available at 
\url{http://www.gtc.iac.es/instruments/osiris}.}. The slit was oriented along the line joining both 
quasar images, and its width was $1\farcs 23$ ($\sim$5 pixel). We took a single 1800 s 
exposure with the blue grism (R500B) on 2015 April 15 under good observing conditions: dark and 
photometric night, FWHM = $0\farcs 95$ at the central wavelength $\lambda_{\rm{c}}$ = 4745 \AA, and 
airmass of 1.04. We also observed the lens system with the red grism (R500R) on 2015 April 16. This 
second grey night, 3 $\times$ 1800 s exposures were taken under variable seeing (FWHM = $1\farcs 
19$, $0\farcs 89$, and $0\farcs 79$ at $\lambda_{\rm{c}}$ = 7165 \AA) and airmass (1.04, 1.06, and 
1.10). Hence, the new data cover a wavelength range of 3570$-$9250 \AA\ with a resolving power of 
$\sim$ 300$-$400. Observations of the spectrophotometric standard star Ross 640 
\citep{1974ApJS...27...21O} were performed with a wider slit of $2\farcs 52$, which forced us to 
carry out a careful flux calibration from photometric data (see below).

For each grism, through a standard data reduction with IRAF, we obtained a sky--subtracted, 
wavelength--calibrated spectrum for each $\sim$5 pixel slice along the slit. However, the 
extraction of the individual spectra of A, B, and G1 is not a simple task. We deal with three 
sources next to each other. Moreover, there is high brightness contrast between B and G1 (e.g., 
$B/G1 \sim$ 40 in the $r$ band), only $0\farcs 4$ apart. These complications prevented detection of 
G1 using the blue grism data. Among the known extraction techniques 
\citep[e.g.,][]{1998AJ....115..377F,2006ApJ...641...70O,2007A&A...468..885S,2014A&A...568A.116S}, 
after some tests, we chose a method similar to that of \citet{2007A&A...468..885S}. They derived 
spectra for three close point--like sources along a slit by fitting three 1D Moffat profiles for 
each wavelength bin. Although G1 is an extended source, it has a very large ellipticity with 
position angle almost perpendicular to the slit axis, and its effective radius in the spatial 
direction is $\sim 0\farcs 13$ \citep{2016MNRAS.458....2R}. Thus, in the spatial direction, this 
very faint galaxy can be treated as a point--like object (we checked the validity of this 
approach). For the red grism, our initial model consisted of three 1D Moffat profiles (A+B+G1) at 
each wavelength, while we only considered two components (A+B) when analysing the blue grism data.

\begin{figure}[ht!]
	\begin{minipage}[h]{0.5\linewidth}
      \centering
      \includegraphics[width=1.05\textwidth]{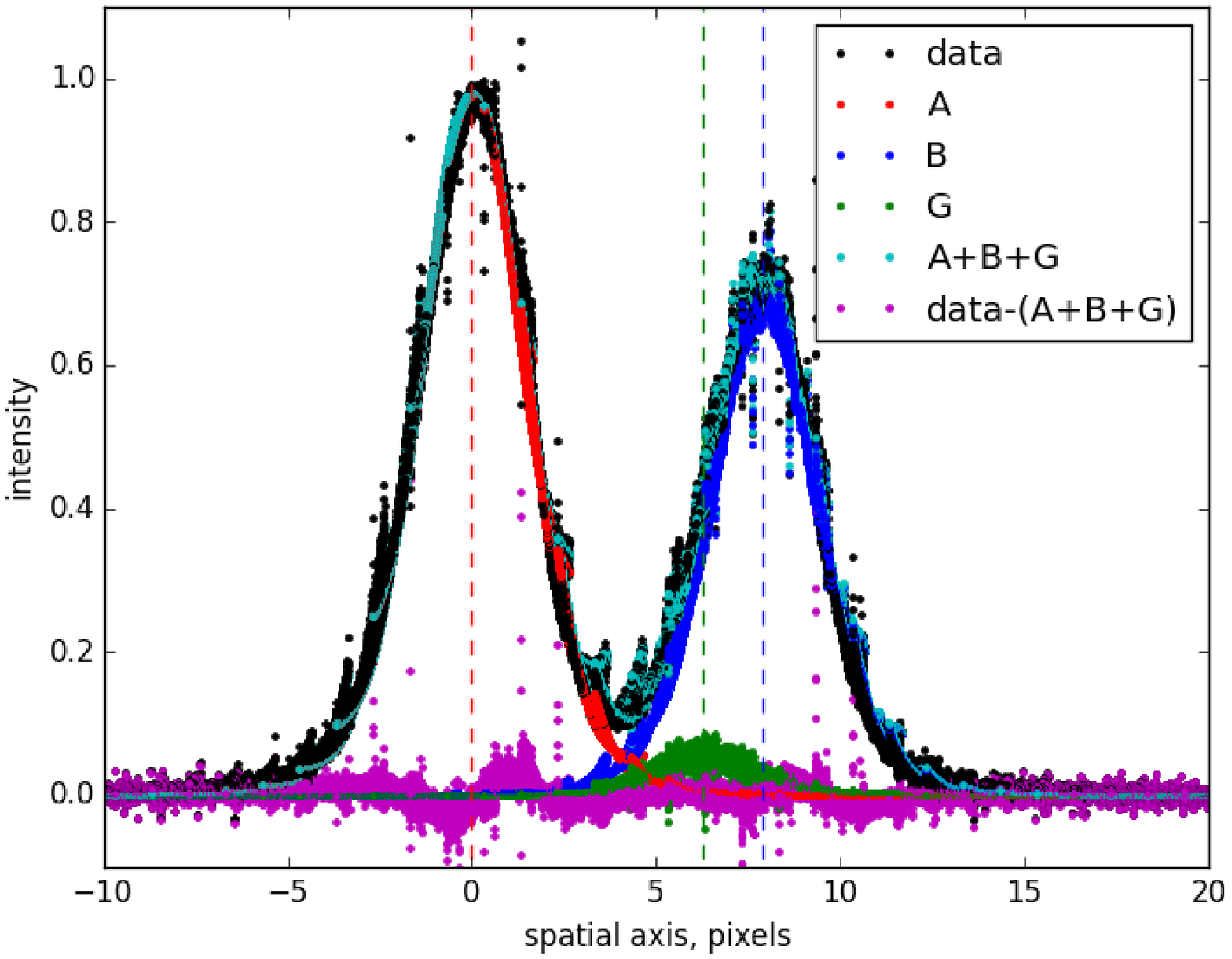}
      \end{minipage}
	\begin{minipage}[h]{0.5\linewidth}
      \centering
      \includegraphics[width=1.05\textwidth]{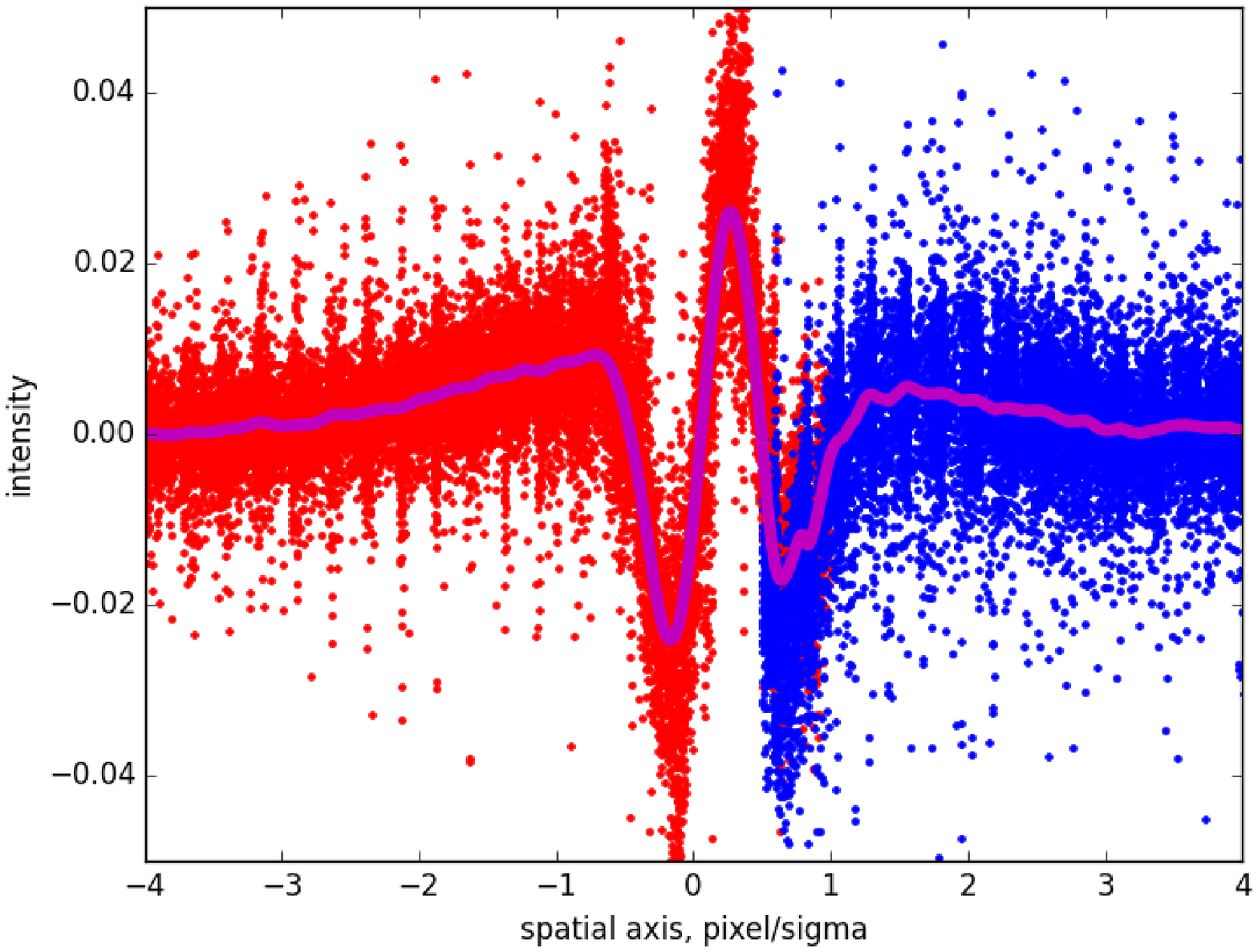}
    	\end{minipage}
\caption{GTC--OSIRIS--R500R fluxes along the slit. Left: Normalized spatial profile as a result of 
combining the 1D flux distributions for all wavelength bins. The measured fluxes (data) are 
compared with fitted fluxes, using a model that consists of three 1D Moffat profiles (A+B+G1) at 
each wavelength. Right: Differences between the true PSF and the Moffat function. These residuals 
(PSF $-$ Moffat) come from the central region and the left wing of A (red circles), and the right 
wing of B (blue circles). The smoothed behaviour (purple line) is the LUT correction (see main 
text).}
\label{fig:f2} 
\end{figure}

From now on, we are going to describe in detail the extraction of the individual spectra of A, B, 
and G1, via the red grism data. We used the Rusu et al.'s astrometry to set the positions of B and 
G1 with respect to A. In addition, the 1D Moffat function is characterized by three parameters: 
centroid, slope, and width, and after preliminary data analysis, we also fixed a global value (the 
same at all wavelengths) for the slope \citep{2007MNRAS.378...23J}. Thus, in a first iteration, 
only the position (centroid) of A, the width of the Moffat function, and the amplitudes of the 
three components were allowed to vary at each wavelength. The first fits to the multi--wavelength 
1D flux distributions generated wavelength--dependent values of the position of A and the width of 
the 1D PSF model, which were fitted to smooth polynomial functions. In a second iteration, these 
two position--structure parameters were evaluated by their polynomial laws, leaving only the 
amplitudes as free parameters. We then obtained the fluxes of A, B, and G1 through the fitted 
amplitudes and the observational priors for the structure parameters of the PSF model.

Although the second iteration produces accurate spectra of the quasar images, the spectrum of G1 is
not as smooth as would be desired. The PSF shape is not perfectly described with a Moffat function,
and residuals (data $-$ model) are comparable to the 1D fluxes of G1. Because of the existence of 
this important residuals--induced noise, we refined the PSF model to try to remove/minimize 
residuals with an empirical look--up--table (LUT) correction \citep[e.g.,][]{2000Msngr..99...31M,
2007MNRAS.378...23J}. Following \citet{2000Msngr..99...31M}, the $\sim$1000 measured spatial 
profiles (1D flux distributions) were combined into the single normalized distribution that is 
shown in the left panel of Figure \ref{fig:f2}. The differences between measured and fitted fluxes 
(purple circles) display a regular spatial pattern, and we focused on a region around and to the 
left of the centroid of A, as well as on a region to the right of the centroid of B. Both regions 
basically include the residuals of interest, i.e., differences between the true PSF and the Moffat 
function. In the right panel of Figure \ref{fig:f2}, we present properly scaled residuals for a PSF 
centered at $x$ = 0. These were smoothed in order to reduce noise (see the purple line in the right 
panel of Figure \ref{fig:f2}), and then incorporated into a refined PSF model: analytical Moffat 
function plus empirical LUT. 

\begin{figure}[ht!]
\begin{minipage}[h]{1.0\linewidth}
	\centering
	\includegraphics[width=0.7\textwidth]{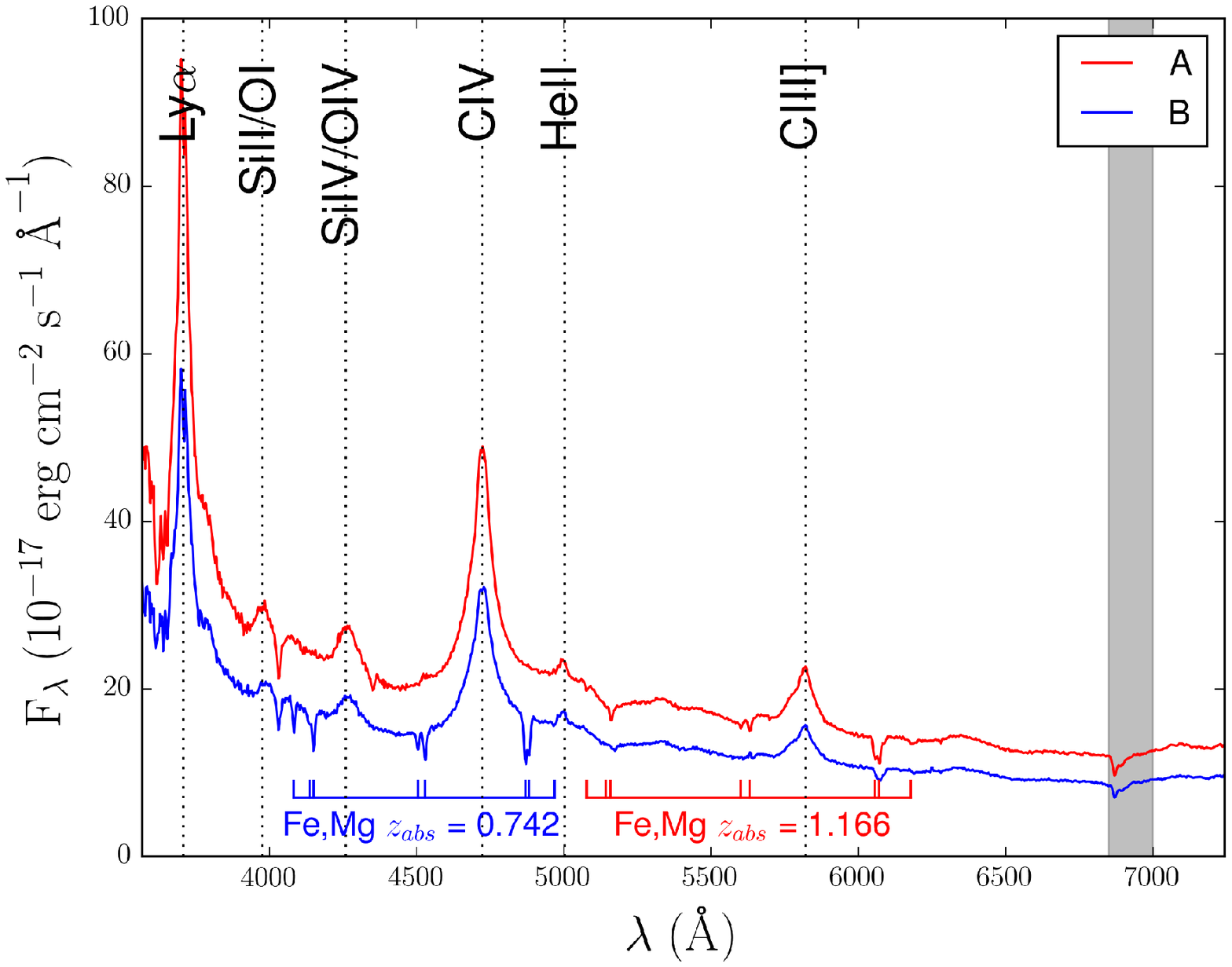}
     	\end{minipage}

\vspace{5.00mm}

	\begin{minipage}[h]{1.0\linewidth}
      \centering
      \includegraphics[width=0.7\textwidth]{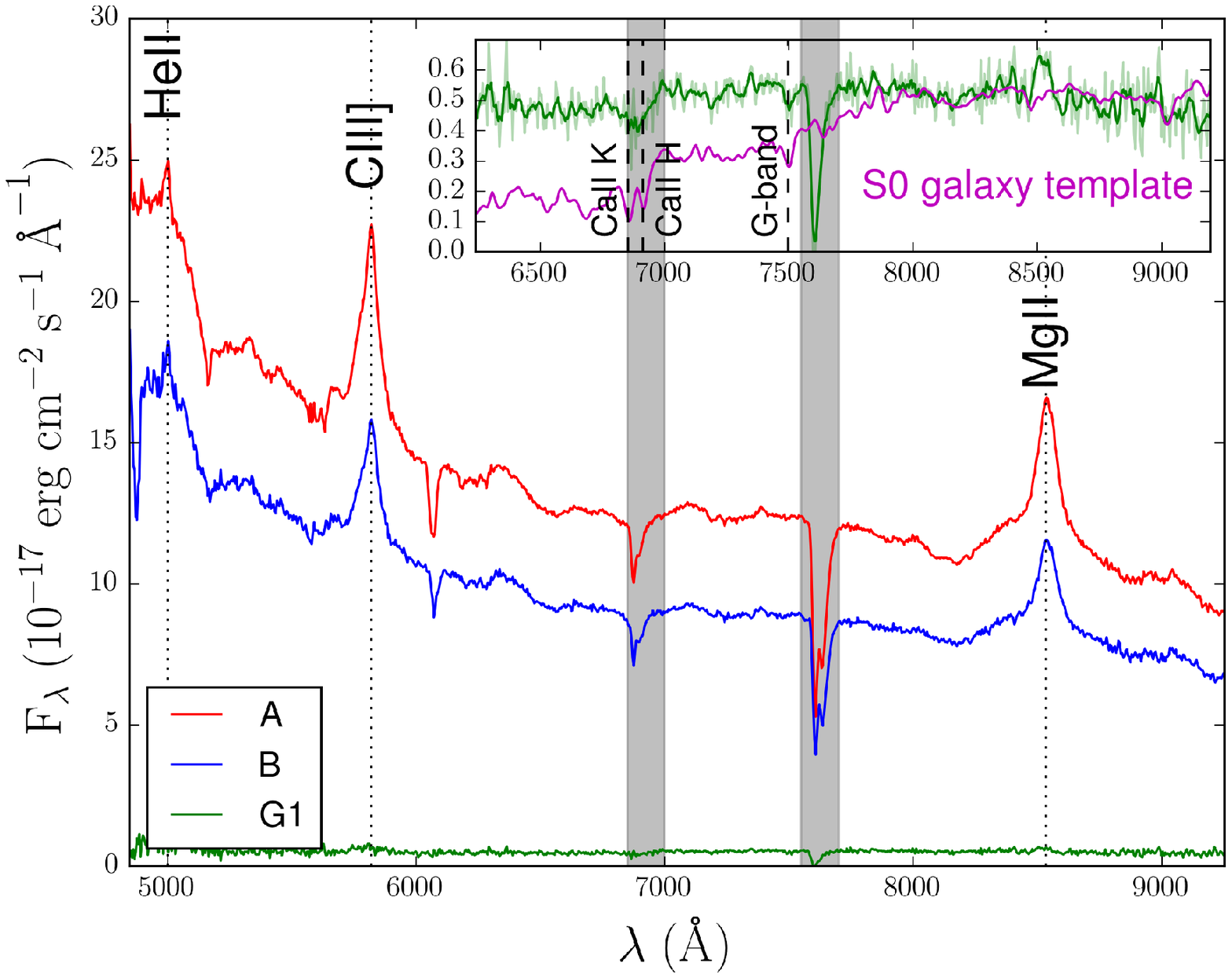}
	\end{minipage}
\caption{GTC--OSIRIS spectra of SDSS J1515+1511ABG1 in 2015. Vertical dotted lines indicate 
emission lines at $z_{\rm{s}}$ = 2.049, while grey highlighted regions are associated with 
atmospheric artefacts. Top: R500B grism. The spectrum of B includes Fe\,{\sc i}, Fe\,{\sc ii}, and 
Mg\,{\sc ii} absorption at $z_{\rm{abs}}$ = 0.742, and the spectrum of A contains a notable Fe/Mg
absorption at $z_{\rm{abs}}$ = 1.166 (see Sections \ref{subsec:dm} and \ref{sec:lensmod}). Bottom: 
R500R grism. The top sub--panel displays a zoomed--in version of the G1 spectrum along with 
the red--shifted ($z$ = 0.742) spectral template of an S0 galaxy \citep{1996ApJ...467...38K}.}
\label{fig:f3}
\end{figure}

In a third iteration, we used Moffat+LUT profiles to describe the contributions of the three 
sources (quasar images and main lensing galaxy) at each wavelength, and thus obtained the final 
instrumental spectra of A, B, and G1. From the instrumental spectrum of the standard star, we also 
built the spectral response function, and corrected the instrumental fluxes of the quasar and the
galaxy. Unfortunately, the standard star was observed with a relatively wide slit (see above), 
which did not allow us to accurately calibrate the flux scale. In order to calibrate the quasar 
spectra, we used $r$--band fluxes of A and B from LT frames taken on 2015 April 16 (see Section 
\ref{subsec:livmon}). To fix the flux scale of G1, we also considered the $i$--band flux of the 
galaxy by \citet{2014AJ....147..153I}. Our final flux--calibrated spectra of the lens system (blue 
and red grisms) are shown in Figure \ref{fig:f3}. Tables \ref{tab:t2} and \ref{tab:t3} include the 
spectra of the lens system from the blue and red grism data, respectively. Additionally, all raw 
and reduced frames in FITS format are publicly available at the GTC 
archive\footnote{\url{http://gtc.sdc.cab.inta-csic.es/gtc/index.jsp}}. 

\begin{deluxetable}{ccc}
\tablecaption{GTC--OSIRIS--R500B spectra of SDSS J1515+1511AB.\label{tab:t2}}
\tablenum{2}
\tablewidth{0pt}
\tablehead{
\colhead{$\lambda$\tablenotemark{a}} & 
\colhead{$F_{\lambda}$(A)\tablenotemark{b}} & 
\colhead{$F_{\lambda}$(B)\tablenotemark{b}}
}
\startdata
3567.142 & 48.233 & 32.704 \\
3570.736 & 47.358 & 30.125 \\
3574.330 & 48.826 & 29.310 \\
3577.924 & 46.260 & 28.706 \\
3581.518 & 48.972 & 32.073 \\
\enddata
\tablenotetext{a}{Observed wavelength in \AA.}
\tablenotetext{b}{Flux in 10$^{-17}$ erg cm$^{-2}$ s$^{-1}$ \AA$^{-1}$.}
\tablecomments{Table 2 is published in its entirety in the machine-readable format.
A portion is shown here for guidance regarding its form and content.}
\end{deluxetable}

\begin{deluxetable}{cccc}
\tablecaption{GTC--OSIRIS--R500R spectra of SDSS J1515+1511ABG1.\label{tab:t3}}
\tablenum{3}
\tablewidth{0pt}
\tablehead{
\colhead{$\lambda$\tablenotemark{a}} & 
\colhead{$F_{\lambda}$(A)\tablenotemark{b}} & 
\colhead{$F_{\lambda}$(B)\tablenotemark{b}} &
\colhead{$F_{\lambda}$(G1)\tablenotemark{b}}
}
\startdata
4846.622 & 26.253 & 18.984 & 0.466 \\
4851.437 & 23.692 & 16.945 & 0.551 \\
4856.251 & 23.546 & 15.886 & 0.300 \\
4861.066 & 23.258 & 14.193 & 0.598 \\
4865.880 & 23.847 & 14.016 & 0.268 \\
\enddata
\tablenotetext{a}{Observed wavelength in \AA.}
\tablenotetext{b}{Flux in 10$^{-17}$ erg cm$^{-2}$ s$^{-1}$ \AA$^{-1}$.}
\tablecomments{Table 3 is published in its entirety in the machine-readable format.
A portion is shown here for guidance regarding its form and content.}
\end{deluxetable}

\begin{figure}[ht!]
\centering
\includegraphics[width=0.7\textwidth]{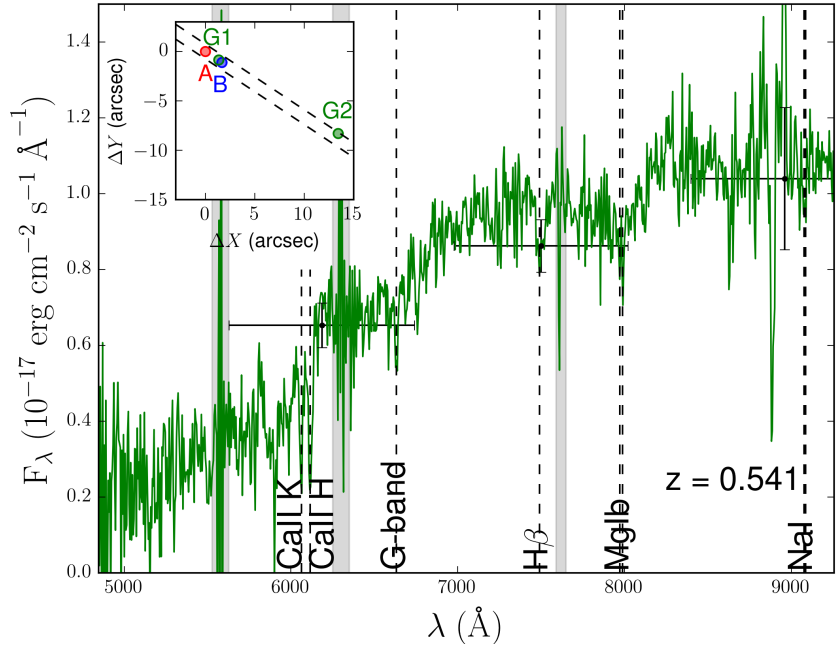}
\caption{GTC--OSIRIS--R500R spectrum of the secondary galaxy G2. We also display several expected 
absorption features at $z_{\rm{G2}}$ = 0.541 (vertical dashed lines), and use crosses and grey 
rectangles to highlight the $riz$--SDSS fluxes and spectral regions associated with atmospheric 
artefacts, respectively. The observational configuration appears in the top sub--panel. This
shows the four sources A, B, G1, and G2 within the slit edges (two parallel dashed lines).}
\label{fig:f4}
\end{figure}

To confirm or not the tentative redshift of the main lensing galaxy 
\citep[0.742;][]{2014AJ....147..153I}, we analysed the G1 spectrum in the top sub--panel inside the 
bottom panel of Figure \ref{fig:f3}. In spite of all our efforts with the GTC data, G1 appears as a 
very faint source with significant contamination by light of B. This contamination distorts its 
spectral shape by producing artefact peaks (e.g., Mg\,{\sc ii} residual flux at $\sim$8500 \AA) and 
an almost flat continuum. Indeed, the spectrum of G1 does not display a noticeable drop off in 
flux at $\sim$7000 \AA, which would correspond to a 4000 \AA\ break in the rest frame of a 
disk--like galaxy without emission lines at a redshift of 0.742 (see the magenta line). Vertical 
dashed lines indicate the positions of some expected absorption features at $z_{\rm{G1}}$ = 0.742, 
but unfortunately, these positions coincide with flux decrements that are indistinguishable from 
noise or are located in a sky absorption region. As a consequence, we were not successful in an 
unambiguous spectroscopic determination of $z_{\rm{G1}}$. However, we were lucky because our 
GTC--OSIRIS--R500R long--slit exposures included light from the secondary galaxy G2 (see Section 
\ref{sec:intro} and the top sub--panel inside Figure \ref{fig:f4}), and thus, we extracted 
the G2 spectrum and measured its redshift. Since G2 is $\sim 16\arcsec$ away from the A image, we 
used the IRAF/APEXRACT package to get an initial spectrum, and then the $riz$--SDSS magnitudes of 
the galaxy to peform an accurate flux calibration. The final spectrum is consistent with an 
early--type galaxy at $z_{\rm{G2}}$ = 0.541 (see Figure \ref{fig:f4}).

\subsubsection{Spectroscopic monitoring} \label{subsubsec:livspec}

We are conducting a robotic monitoring of a small sample of $\sim$10 lensed quasars with images 
brighter than $r$ = 20 mag and visible from the Northern Hemisphere 
\citep{2015arXiv150504317G}. Taking the spatial resolution of the LT spectrographs into account, we 
only consider spectroscopic monitoring campaigns of double quasars with image separation $\geq 
2\arcsec$. For each such doubles, we obtain spectroscopic observations separated by the time delay 
between its two images, which allows us to study magnification ratios for the continuum and the 
emission lines at different wavelengths. These delay--corrected ratios (i.e., at the same emission 
time) are key tools to discuss the differential dust extinction, the macrolens magnification ratio, 
and microlensing effects in the system \citep[e.g.,][]{2006glsw.conf.....M}.
     
We observed \object{SDSS J1515+1511} with the SPRAT long--slit spectrograph on 2015 August 16. The 
$1\farcs 8$ ($\sim$4 pixel) wide slit was oriented along the line joining A and B, and we used the
red grating mode. The grating may be set to two different configurations which are optimized for 
the blue or red regions of the 4000$-$8000 \AA\ wavelength range (resolving power of $\sim$350 at
6000 \AA). This first night, we took 5 $\times$ 600 s exposures under normal observing conditions: 
FWHM $\sim 1\farcs 5$ at 6000 \AA\ and airmass of $\sim$1.3. We also obtained 5 $\times$ 600 s 
LT--SPRAT--blue exposures on 2015 August 18 to check the best grating mode for this lens system. 
After getting the earliest SPRAT spectra in 2015 August, a second step was to decide on a sampling 
time based on some time delay estimation. While current lens models predict delays ranging from 5 
to 7 months \citep{2016MNRAS.458....2R}, we used our two first photometric monitoring seasons (see
Sections \ref{subsec:livmon} and \ref{sec:delay}) to set the sampling time to 7 months. Thus, we
have re--observed the system on 2016 March 17 to obtain 5 $\times$ 600 s LT--SPRAT--blue exposures 
under acceptable seeing conditions (FWHM $\sim 1\farcs 7$ at 6000 \AA) with a low airmass of 
$\sim$1.05. We also observed the spectrophotometric standard star BD+33d2642 
\citep{1990AJ.....99.1621O} on each of the three monitoring nights in 2015$-$2016. 

\begin{figure}[ht!]
	\begin{minipage}[h]{0.5\linewidth}
      \centering
      \includegraphics[width=1.05\textwidth]{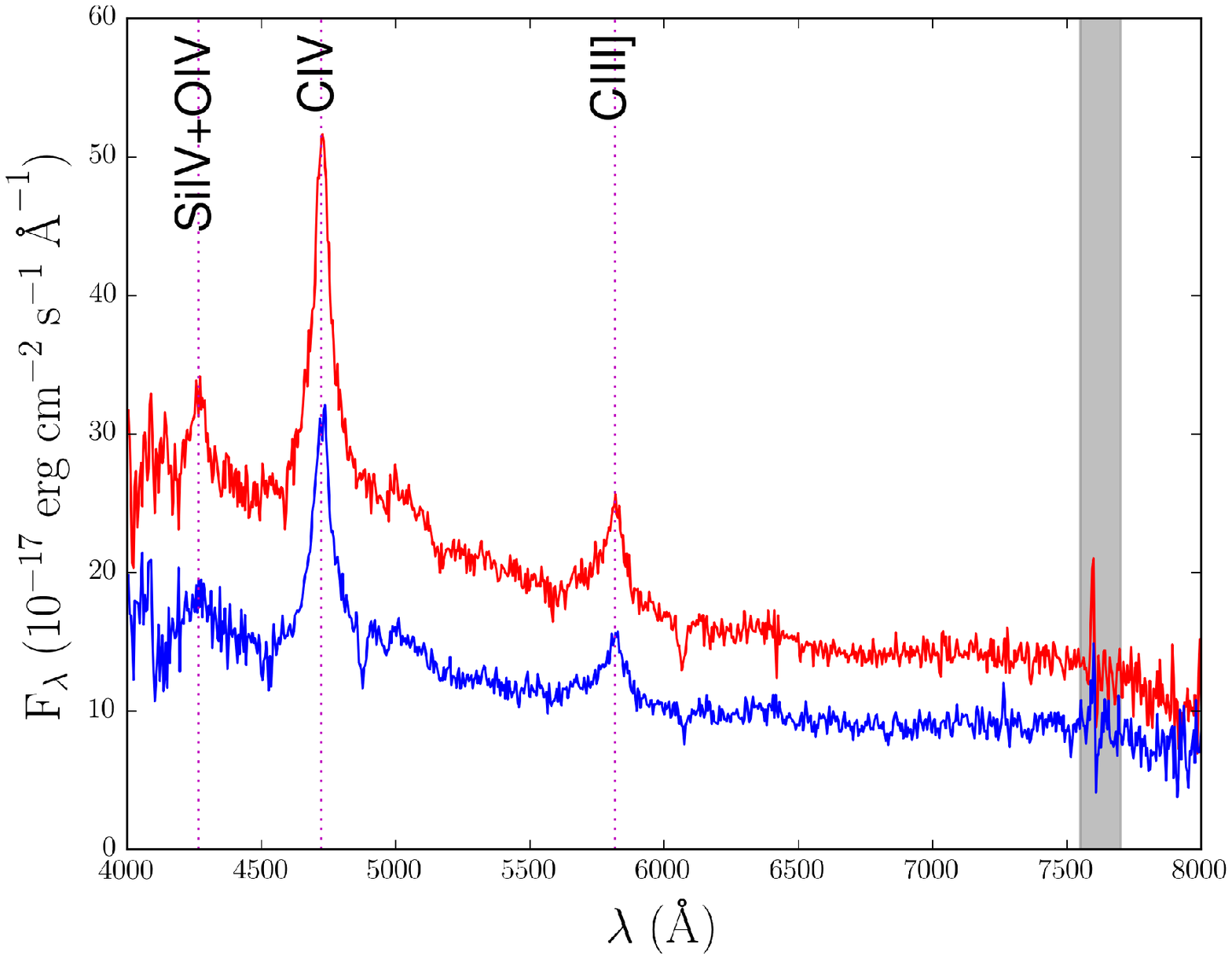}
      \end{minipage}
	\begin{minipage}[h]{0.5\linewidth}
      \centering
      \includegraphics[width=1.05\textwidth]{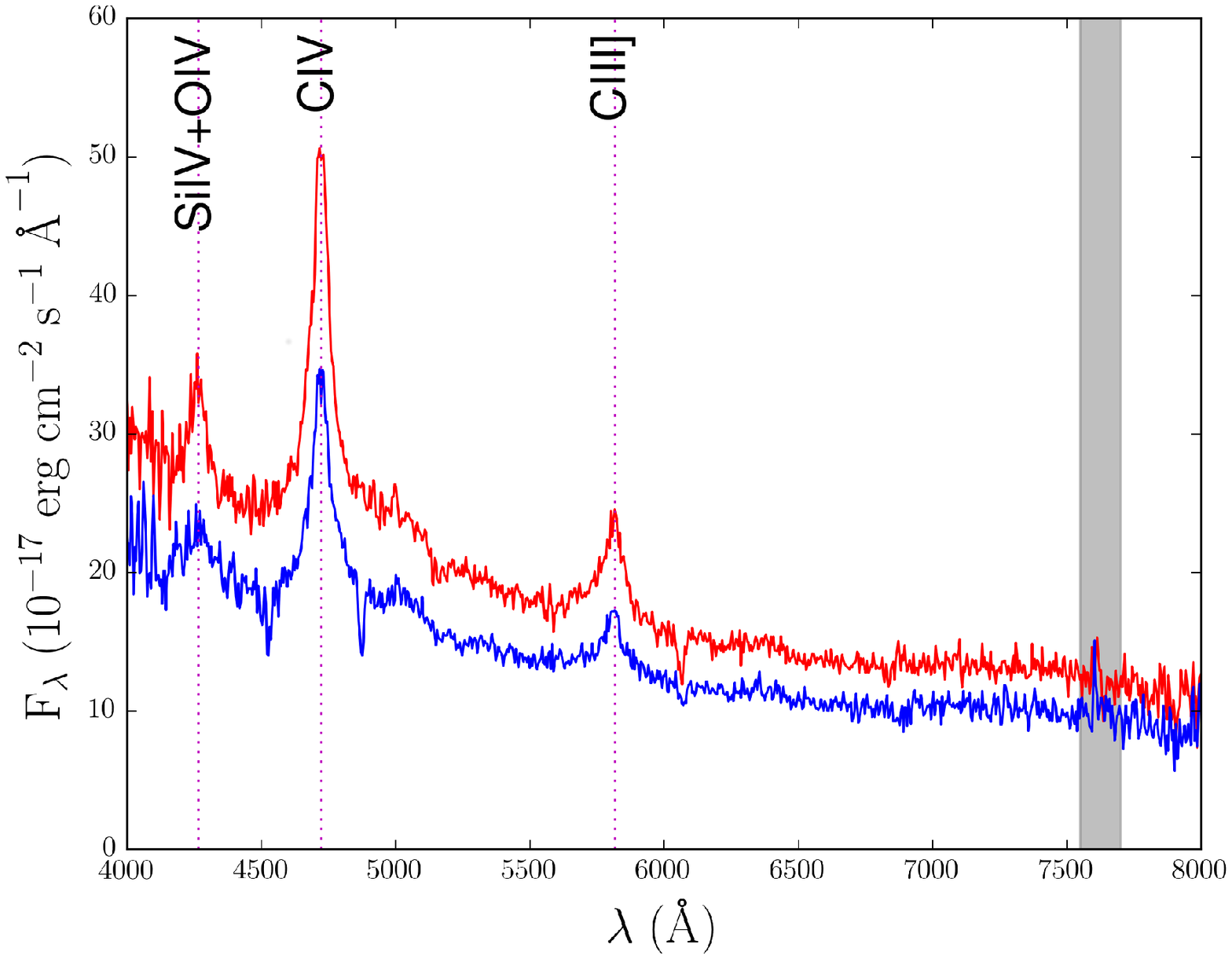}
    	\end{minipage}
\caption{LT--SPRAT spectra of SDSS J1515+1511AB. Vertical dotted lines indicate emission lines at 
$z_{\rm{s}}$ = 2.049, while grey highlighted regions are associated with atmospheric artefacts. 
Left: 2015 August (total exposure time = 6 ks). Right: 2016 March (total exposure time = 3 ks).}
\label{fig:f5} 
\end{figure}

The standard data reduction with IRAF included the same tasks that we used to process the GTC 
observations in Section \ref{subsubsec:gtcspec}. The spectra of the two quasar images were 
extracted by fitting two 1D Gaussian profiles with fixed separation. In a first iteration, the 
position of A, the width of the 1D PSF model, and the amplitudes of the two components were allowed 
to vary at each wavelength. In a second iteration, we only considered the multi--wavelength 
instrumental fluxes of A and B as free parameters (see Section \ref{subsubsec:gtcspec}). From these 
instrumental spectra, the spectral response functions, and $r$--band fluxes from LT frames (taken 
on 2015 August 19 and 2016 March 18), we derived flux--calibrated spectra of A and B at the three 
observing epochs. The quasar spectra at the two first close epochs were then combined to make two 
spectral energy distributions (A and B) with lower noise (see the left panel of Figure 
\ref{fig:f5}). All final spectra are shown in Figure \ref{fig:f5}, and are also available in a 
tabular format (Tables \ref{tab:t4} and \ref{tab:t5}). These contain Si\,{\sc iv}+O\,{\sc iv}], 
C\,{\sc iv}, and C\,{\sc iii}] emission lines. It is also evident that the B spectrum varied 
appreciably on a timescale equal to the time delay of the lens system (see Section 
\ref{sec:delay}). 

\begin{deluxetable}{ccc}
\tablecaption{LT--SPRAT--blue/red spectra of SDSS J1515+1511AB in 2015 August.\label{tab:t4}}
\tablenum{4}
\tablewidth{0pt}
\tablehead{
\colhead{$\lambda$\tablenotemark{a}} & 
\colhead{$F_{\lambda}$(A)\tablenotemark{b}} & 
\colhead{$F_{\lambda}$(B)\tablenotemark{b}}
}
\startdata
3980.459 & 38.616 & 13.170 \\
3985.095 & 37.200 & 16.612 \\
3989.731 & 30.862 & 20.126 \\
3994.368 & 32.590 & 18.687 \\
3999.004 & 35.608 & 17.765 \\
\enddata
\tablenotetext{a}{Observed wavelength in \AA.}
\tablenotetext{b}{Flux in 10$^{-17}$ erg cm$^{-2}$ s$^{-1}$ \AA$^{-1}$.}
\tablecomments{Table 4 is published in its entirety in the machine-readable format.
A portion is shown here for guidance regarding its form and content.}
\end{deluxetable}

\begin{deluxetable}{ccc}
\tablecaption{LT--SPRAT--blue spectra of SDSS J1515+1511AB in 2016 March.\label{tab:t5}}
\tablenum{5}
\tablewidth{0pt}
\tablehead{
\colhead{$\lambda$\tablenotemark{a}} & 
\colhead{$F_{\lambda}$(A)\tablenotemark{b}} & 
\colhead{$F_{\lambda}$(B)\tablenotemark{b}} 
}
\startdata
3968.467 & 25.244 & 24.423 \\
3973.122 & 35.824 & 22.400 \\
3977.776 & 40.208 & 24.188 \\
3982.431 & 36.540 & 25.976 \\
3987.086 & 34.417 & 26.696 \\
\enddata
\tablenotetext{a}{Observed wavelength in \AA.}
\tablenotetext{b}{Flux in 10$^{-17}$ erg cm$^{-2}$ s$^{-1}$ \AA$^{-1}$.}
\tablecomments{Table 5 is published in its entirety in the machine-readable format.
A portion is shown here for guidance regarding its form and content.}
\end{deluxetable}

\section{Time delay} \label{sec:delay}

\begin{figure}[ht!]
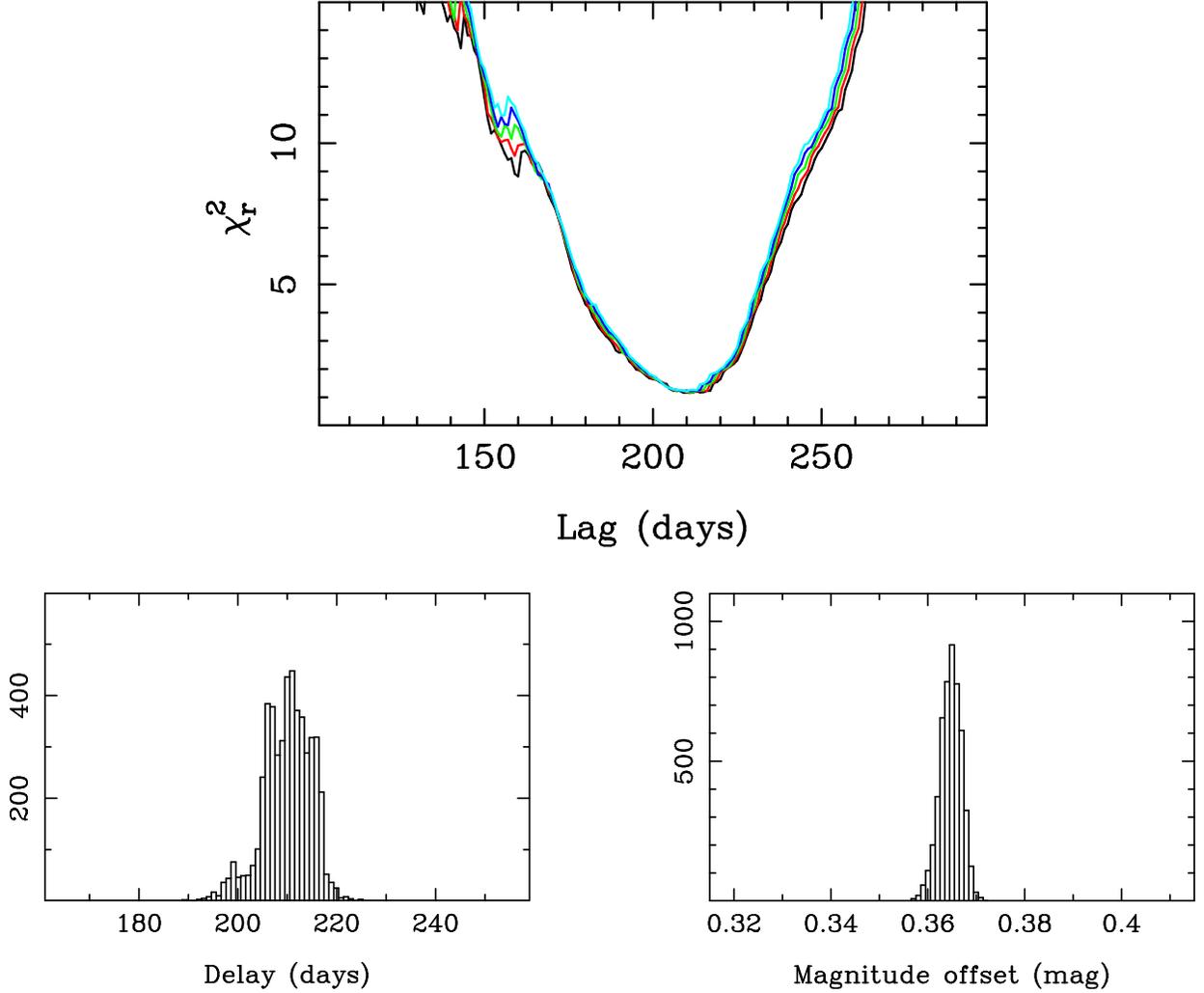

	\begin{minipage}[h]{1.0\linewidth}
	\centering
	\includegraphics[angle=-90,width=0.6\textwidth]{Chi2spec.eps}
     	\end{minipage}

\vspace{5.00mm}

	\begin{minipage}[h]{0.5\linewidth}
      \centering
      \includegraphics[angle=-90,width=0.8\textwidth]{Chi2hdel.eps}
      \end{minipage}
	\begin{minipage}[h]{0.5\linewidth}
      \centering
      \includegraphics[angle=-90,width=0.8\textwidth]{Chi2hoff.eps}
    	\end{minipage}
\caption{Top: $\chi^2_{\rm{r}}$--lag relationships for $\alpha$ = 8 days (black), $\alpha$ = 9 days 
(red), $\alpha$ = 10 days (green), $\alpha$ = 11 days (blue), and $\alpha$ = 12 days (cyan). 
Bottom: Histograms from 1000 pairs of synthetic curves and the $\chi^2_{\rm{r}}$ minimization 
($\alpha$ = 8$-$12 days). The left and right panels display the best solutions of the time delay 
and the magnitude offset, respectively.}
\label{fig:f6} 
\end{figure}

Assuming that the magnitude fluctuations in the right panel of Figure \ref{fig:f1} mainly arise 
from intrinsic variations in the source quasar, we focused on two standard techniques to measure 
the time delay between A and B. First, we carried out a reduced chi--square ($\chi^2_{\rm{r}}$) 
minimization. This $\chi^2_{\rm{r}}$ minimization was based on a comparison between the light curve 
of A and the time--shifted light curve of B, using different lags and bins with semisize $\alpha$ 
in B \citep[e.g.,][]{2006A&A...452...25U}. For $\alpha \sim$ 10 days, we found best solutions 
$\Delta t_{\rm{AB}} \sim$ 210 days ($\chi^2_{\rm{r}} \sim$ 1), which are related with deep minima 
in $\chi^2_{\rm{r}}$--lag relationships (see the top panel of Figure \ref{fig:f6}). The 
uncertainties in $\Delta t_{\rm{AB}}$ and the $r$--band magnitude offset ($\Delta r_{\rm{AB}}$) 
were derived from 1000 repetitions of the experiment (pairs of synthetic light curves based on the 
observed records). To obtain synthetic curves for A and B, we modified the observed magnitudes by 
adding random quantities. These random quantities were realizations of normal distributions around 
zero, with standard deviations equal to the measured errors. We obtained 5000 delays and magnitude 
offsets by applying the $\chi^2_{\rm{r}}$ minimization ($\alpha$ = 8, 9, 10, 11 and 12 days) to the 
1000 pairs of synthetic curves, and the corresponding distributions are shown in the bottom panels 
of Figure \ref{fig:f6}. 

From the histograms in the bottom panels of Figure \ref{fig:f6}, we inferred the 1$\sigma$ 
measurements (68\% confidence intervals) in the first result row of Table \ref{tab:t6}. The time 
delay is 211$^{+4}_{-5}$ days (A is leading), while the $r$--band magnitude offset (magnification 
ratio) is 0.3650 $\pm$ 0.0025 mag. We also used the dispersion minimization to estimate the delay 
and the offset. Thus, the $D^2_{4,2}$ estimator \citep{1996A&A...305...97P} with decorrelation 
lengths $\delta$ = 8$-$12 days produced a minimum at 211 days. Additionally, when applying the 
$D^2_{4,2}$ minimization ($\delta$ = 8$-$12 days) to the 1000 pairs of synthetic curves, we 
obtained the second result row of Table \ref{tab:t6}. Table \ref{tab:t6} also provides composite 
measures ($\chi^2_{\rm{r}} + D^2_{4,2}$) of the time delay and the magnitude offset, which have
2.4\% and 0.7\% accuracy, respectively.

\begin{deluxetable}{ccc}
\tablecaption{Time delay and magnitude offset in the $r$ band of SDSS J1515+1511.\label{tab:t6}}
\tablenum{6}
\tablewidth{0pt}
\tablehead{
\colhead{Method} & \colhead{$\Delta t_{\rm{AB}}$} & \colhead{$\Delta r_{\rm{AB}}$}   
}
\startdata
$\chi^2_{\rm{r}}$				&	211$^{+4}_{-5}$ 	&	0.3650 $\pm$ 0.0025\\
$D^2_{4,2}$					&	211$^{+5}_{-4}$ 	&	0.3650 $\pm$ 0.0025\\
$\chi^2_{\rm{r}} + D^2_{4,2}$ 	&	211 $\pm$ 5 	&	0.3650 $\pm$ 0.0025\\	  
\enddata
\tablecomments{$\Delta t_{\rm{AB}}$ in days and $\Delta r_{\rm{AB}}$ in magnitudes. A is leading, 
and all measurements are 68\% confidence intervals.}
\end{deluxetable}

\begin{figure}[ht!]
\centering
\includegraphics[angle=-90,width=0.7\textwidth]{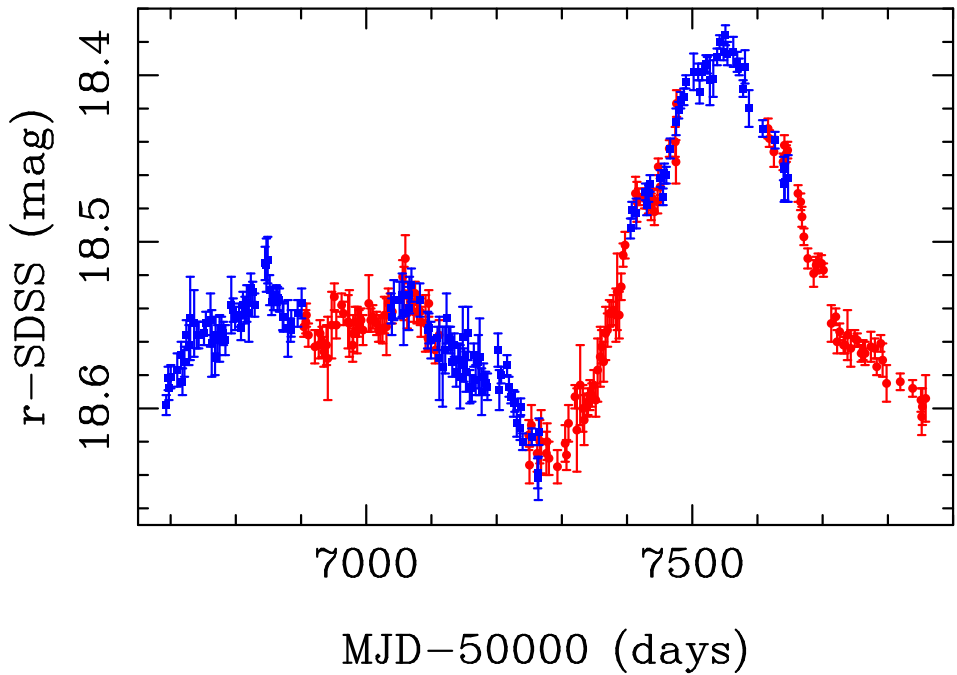}
\caption{Combined light curve in the $r$ band. The B curve (blue squares) and the magnitude-- and 
time--shifted A curve (red circles) are drawn together (see main text).}
\label{fig:f7}
\end{figure}

From the central values in the time delay and magnitude offset intervals (211 days and 0.365 mag), 
we constructed the combined light curve in the $r$ band (see Figure \ref{fig:f7}). Such combined 
curve consists of the B light curve (blue squares) and the magnitude-- and time--shifted brightness 
record of A (red circles). Magnitude fluctuations in both quasar images agree well each other, 
indicating the absence of significant microlensing variability over the 2.6 years (three full
seasons) of monitoring. This is in good agreement with our initial hypothesis in the beginning of 
this section. Despite the existence of sophisticated methods for determining time delays in 
presence of microlensing \citep[e.g.,][]{2013A&A...553A.120T}, simpler standard techniques are 
enough here, and we adopt our composite measures as the final 1$\sigma$ intervals. 

The $\Delta t_{\rm{AB}}$ interval in the last result row of Table \ref{tab:t6} is marginally 
consistent with the delay interval for the SIE+$\gamma$ lens model in Table 5 of 
\citet{2016MNRAS.458....2R}. Hence, the constraints used in that paper seem to be close to the 
actual ones. In Section \ref{sec:extmagmic}, assuming the tentative redshift $z_{\rm{G1}}$ = 0.742 
(see Section \ref{sec:intro}), we discuss the macrolens magnification ratio from the GTC and LT 
spectroscopic data. This is compared with the $K'$--band magnitude difference $B - A$ (at the same 
observing time) in the Rusu et al.'s paper. In Section \ref{sec:lensmod}, the new constraints on 
the macrolens magnification ratio and the time delay are used to check the current SIE+$\gamma$ 
lens model and to validate the assumed redshift of G1 \citep[e.g.,][]{2010ApJ...708..995G}. 

\section{Dust extinction, macrolens magnification and microlensing effect} \label{sec:extmagmic}

\subsection{Analysis of the main emission lines} \label{subsec:mel}

\begin{figure}[ht!]
	\begin{minipage}[h]{0.5\linewidth}
      \centering
      \includegraphics[width=1.0\textwidth]{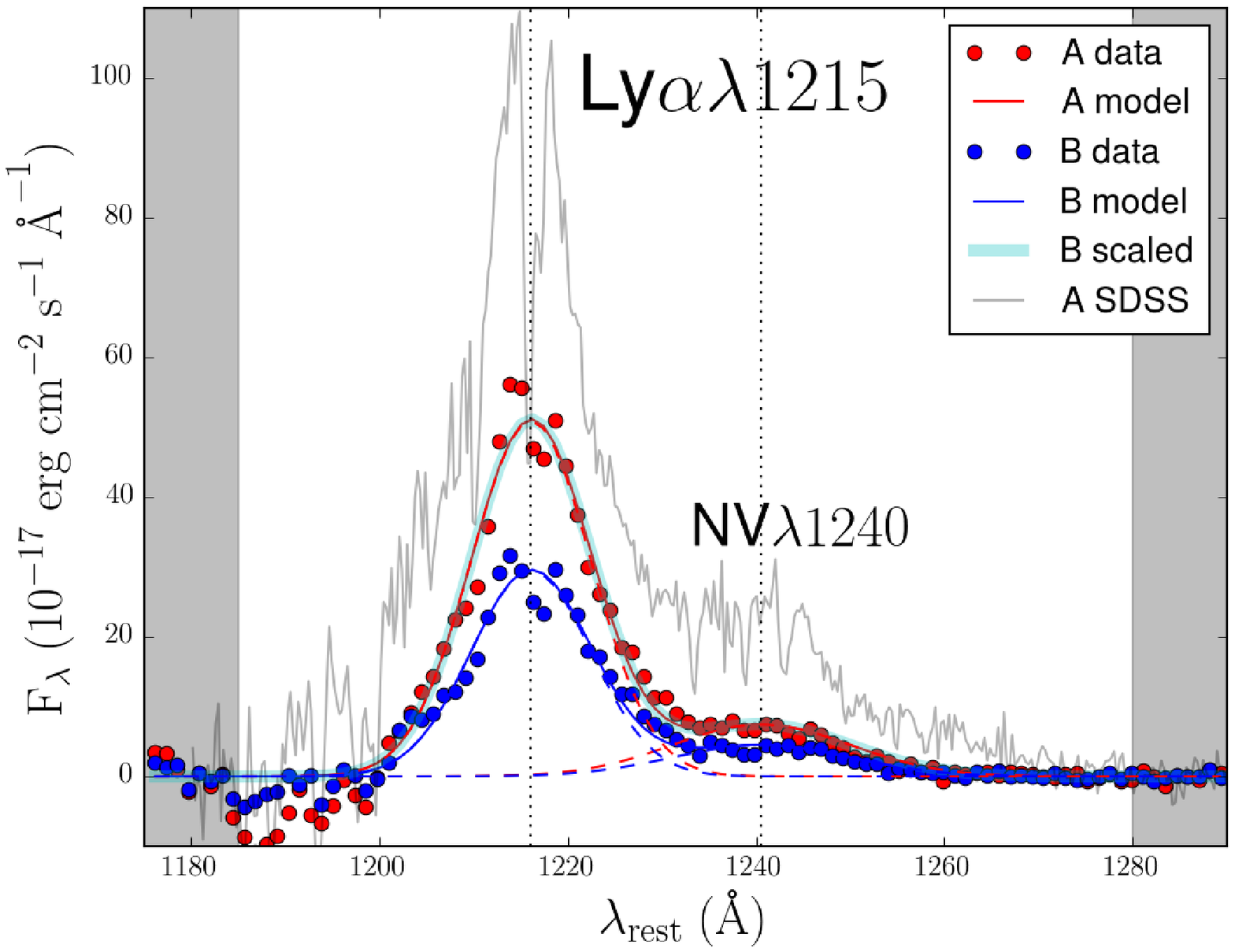}
      \end{minipage}
	\begin{minipage}[h]{0.5\linewidth}
      \centering
      \includegraphics[width=1.0\textwidth]{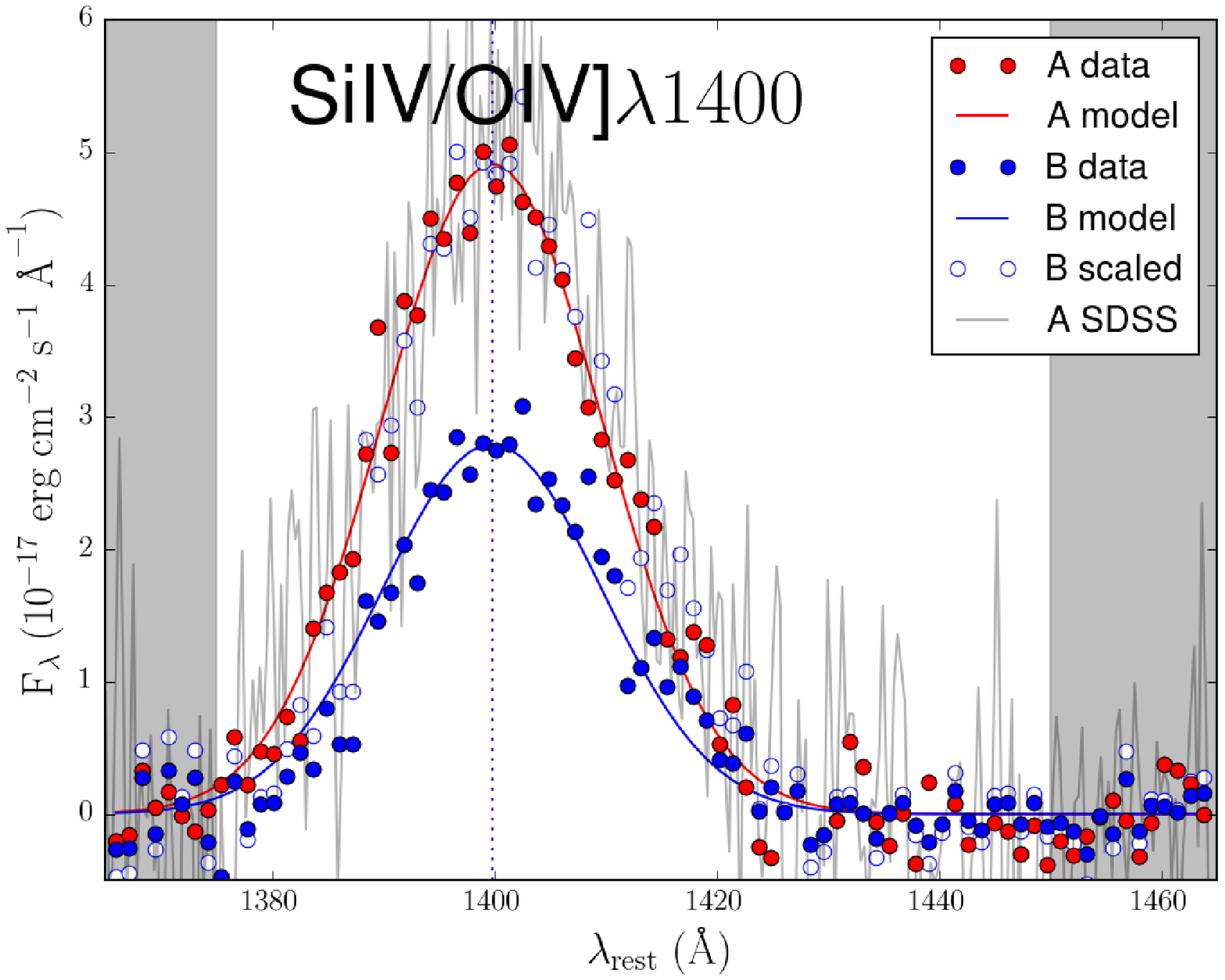}
    	\end{minipage}

\vspace{3.00mm}

	\begin{minipage}[h]{0.5\linewidth}
      \centering
      \includegraphics[width=1.0\textwidth]{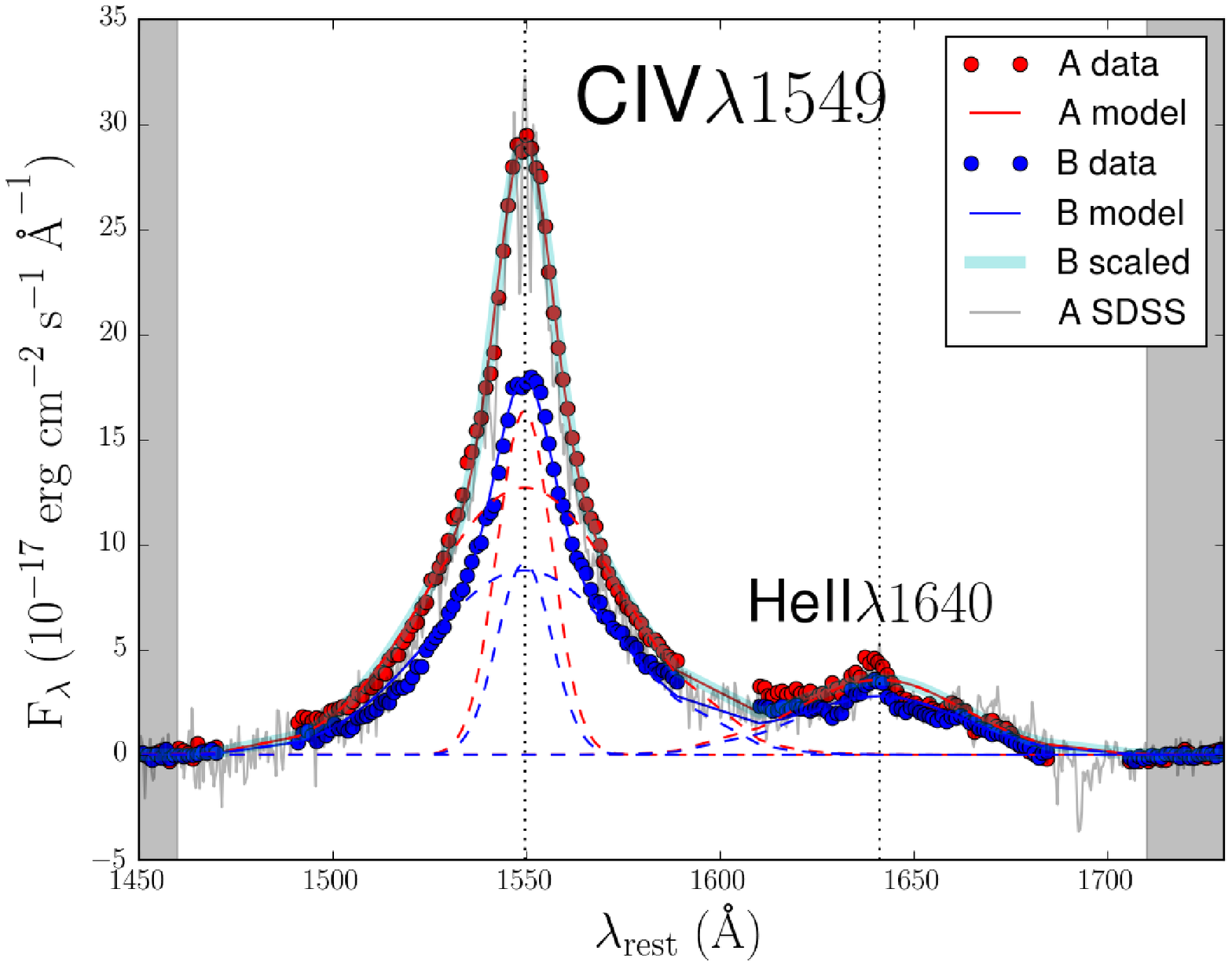}
      \end{minipage}
	\begin{minipage}[h]{0.5\linewidth}
      \centering
      \includegraphics[width=1.0\textwidth]{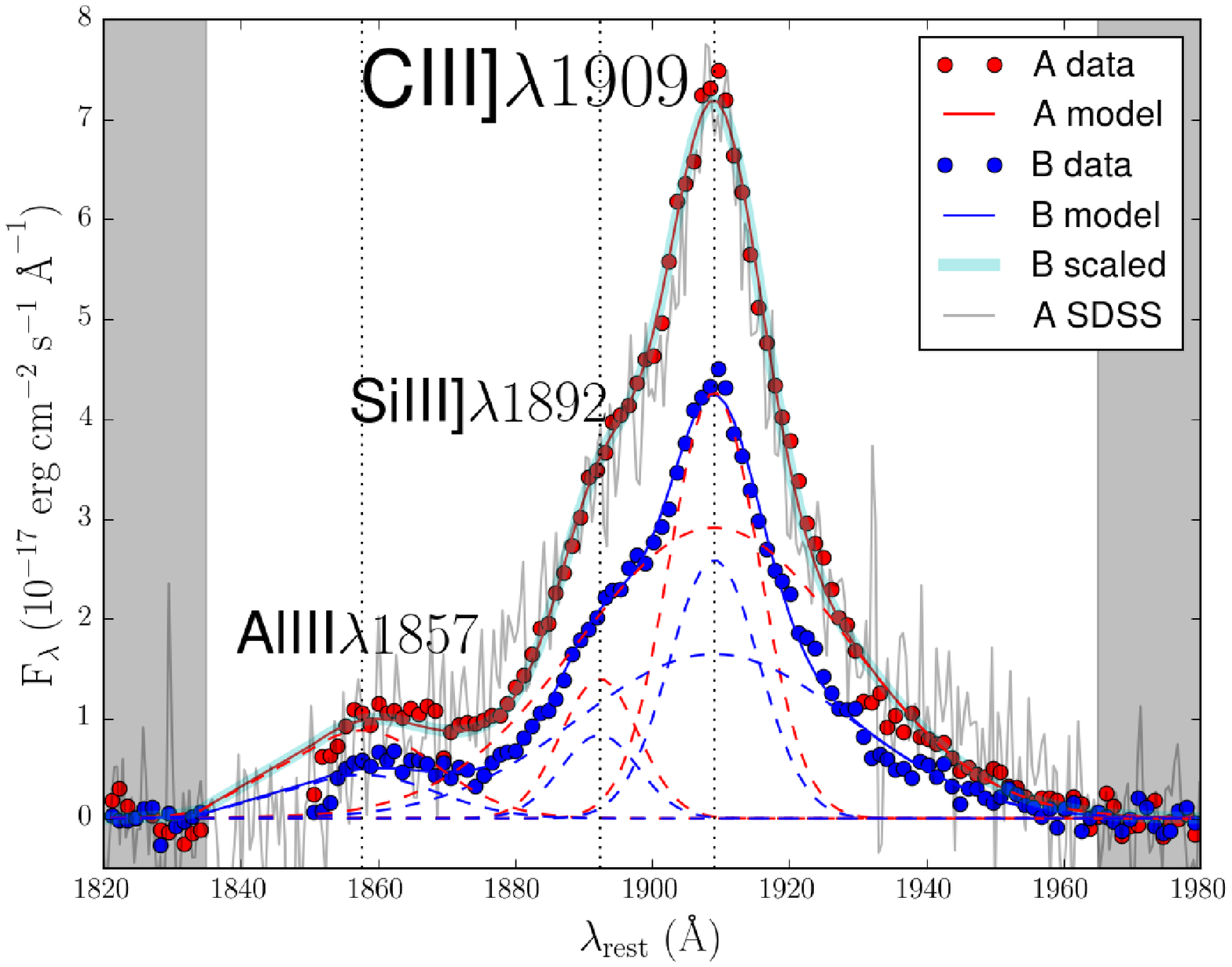}
    	\end{minipage}

\vspace{3.00mm}

	\begin{minipage}[h]{1.0\linewidth}
      \centering
      \includegraphics[width=0.5\textwidth]{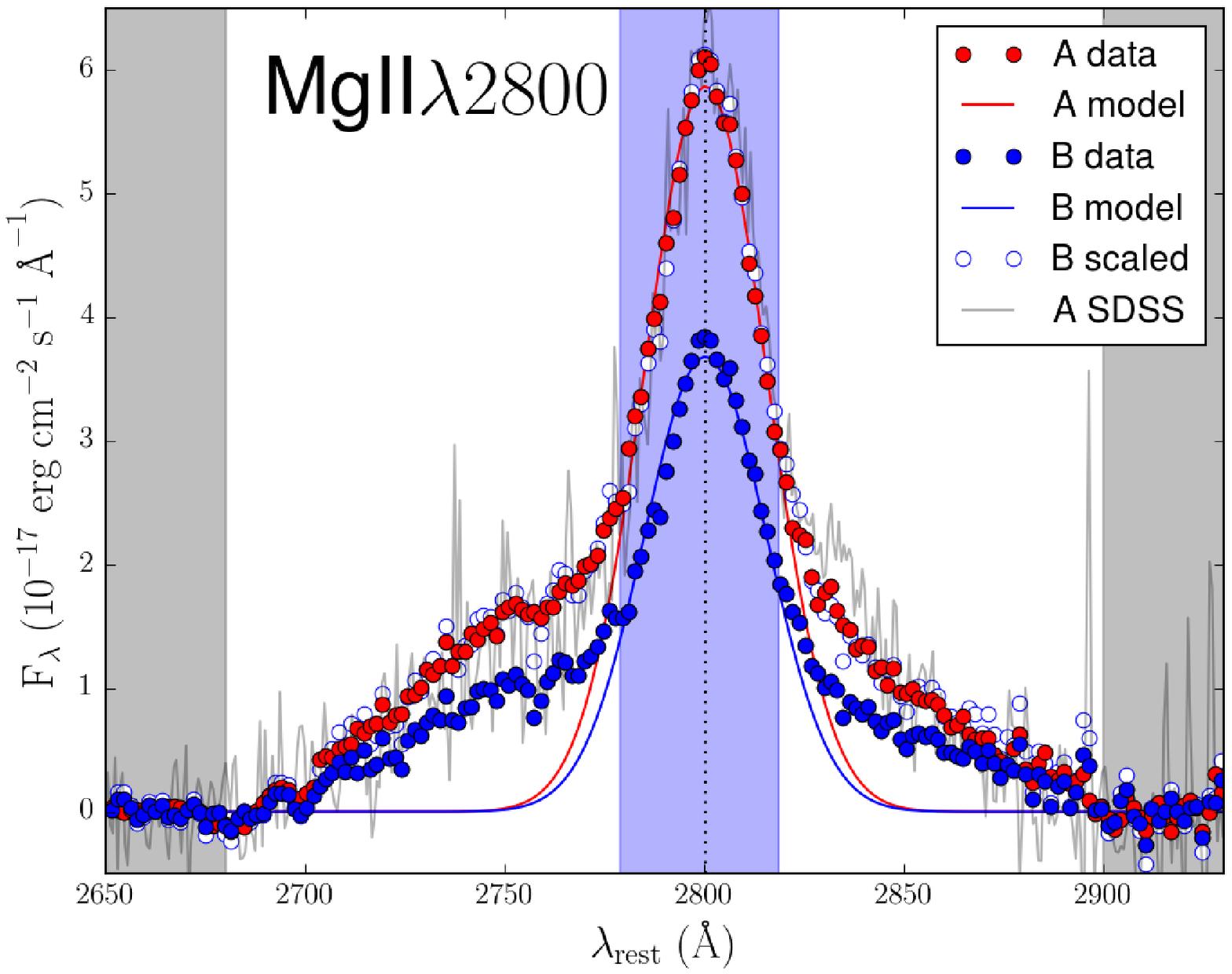}
      \end{minipage}
\caption{Line profiles and multi--component decompositions from the GTC--OSIRIS data of A and B. 
The top and middle panels display the results for the Ly$\alpha$, Si\,{\sc iv}/O\,{\sc iv}], 
C\,{\sc iv}, and C\,{\sc iii}] emission lines in the blue grism spectra, whereas the bottom panel 
shows the results for the Mg\,{\sc ii} emission in the red grism spectra. The profiles for the A 
image on 2015 April 15$-$16 (red circles) are compared with profiles from the SDSS/BOSS spectrum of 
A on 2012 April 13 (grey lines). We also highlight the continuum windows (grey rectangles), as well 
as the 40 \AA\ width central region of the Mg\,{\sc ii} line (blue rectangle in the bottom panel).
See main text for details.}
\label{fig:f8} 
\end{figure}

\begin{figure}[ht!]
\begin{minipage}[h]{0.5\linewidth}
      \centering
      \includegraphics[width=1.0\textwidth]{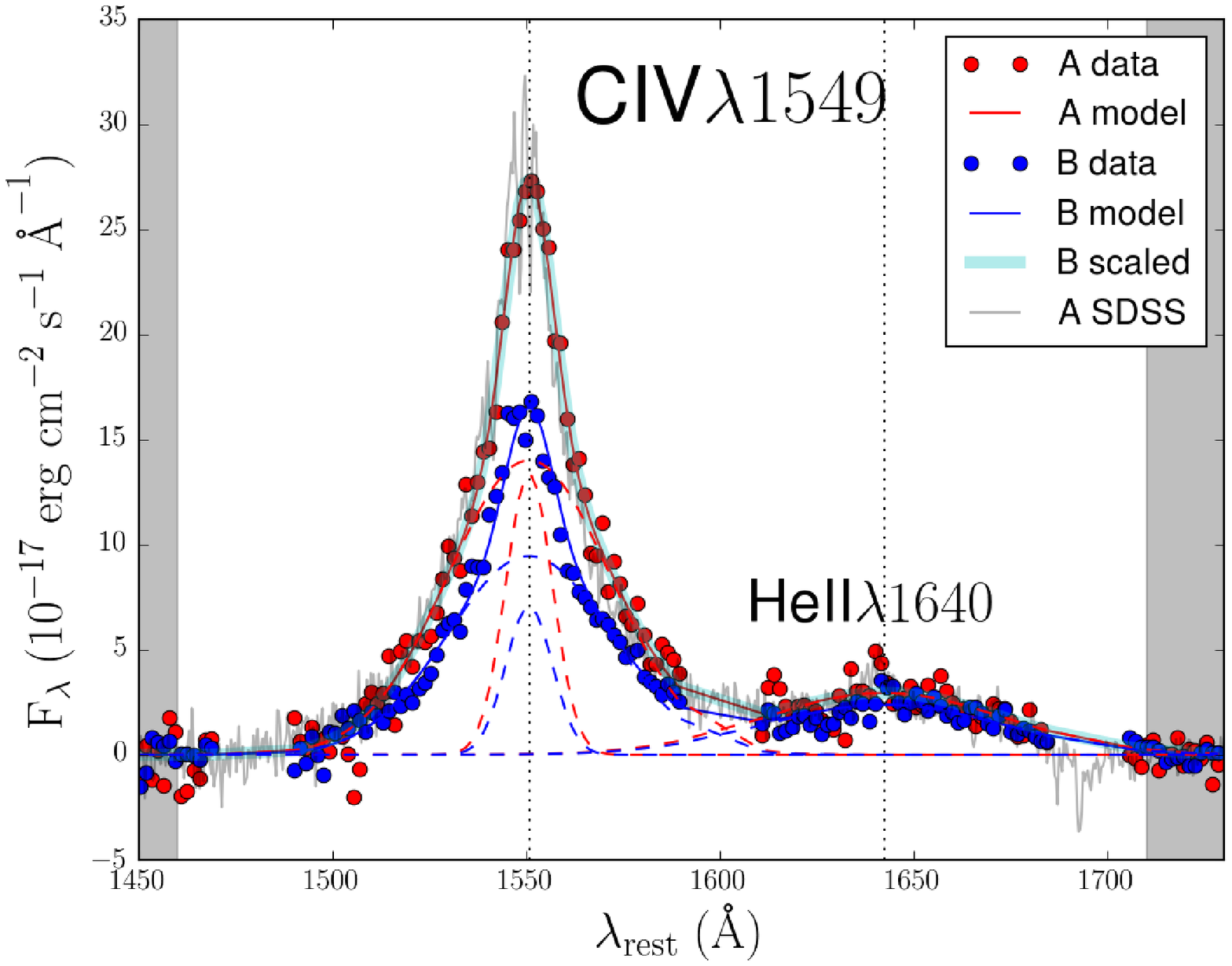}
      \end{minipage}
	\begin{minipage}[h]{0.5\linewidth}
      \centering
      \includegraphics[width=1.0\textwidth]{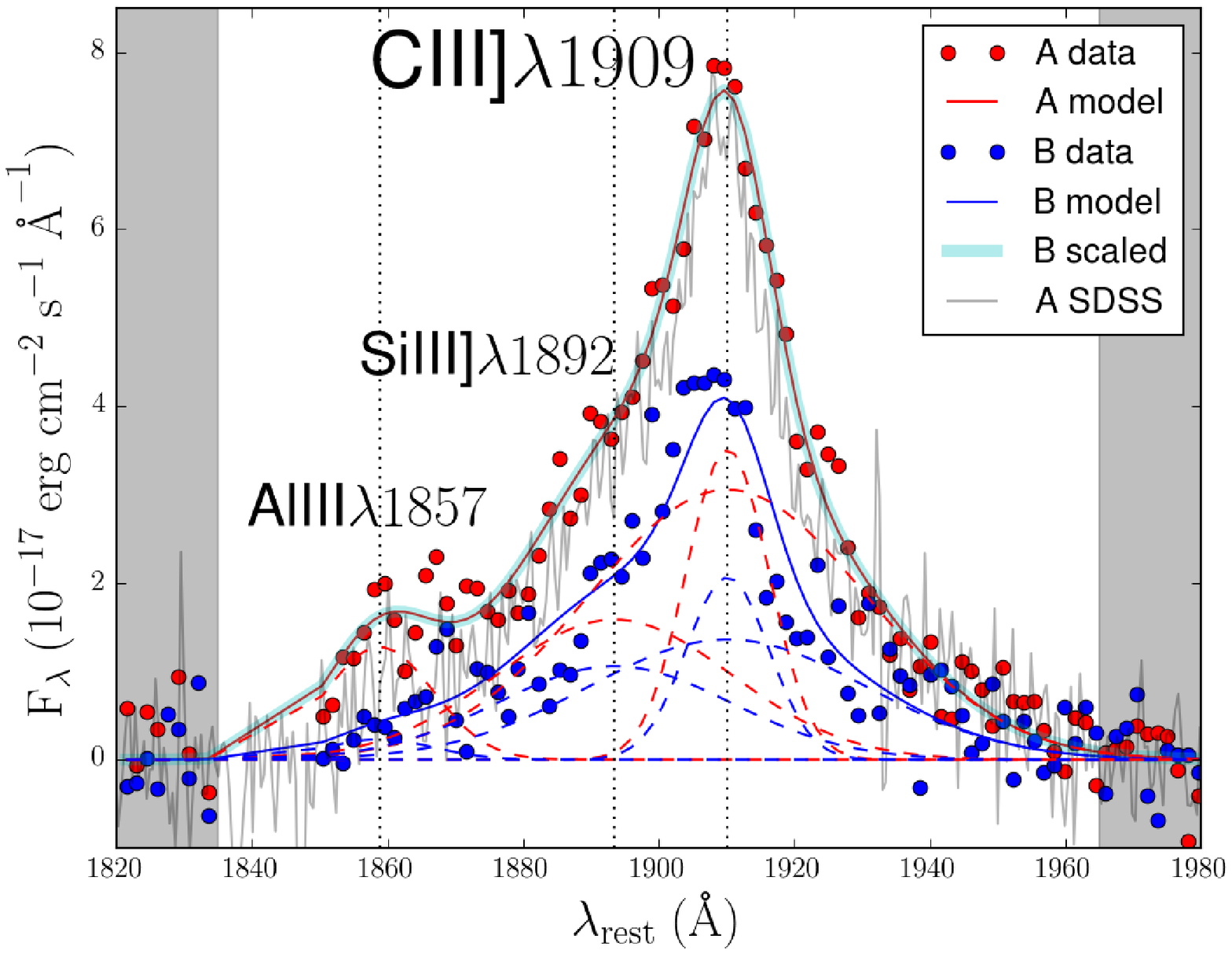}
    	\end{minipage}
\caption{Line profiles and multi--component decompositions from the LT--SPRAT data of A and B. We 
compare the A profiles/decompositions on 2015 August 16$-$18 with the B profiles/decompositions on 
2016 March 17. We also show the continuum windows used to extract the line shapes (grey 
rectangles) and the A profiles from SDSS/BOSS data on 2012 April 13 (grey lines). See main text for 
details.}
\label{fig:f9} 
\end{figure}

The GTC--OSIRIS spectra of SDSS J1515+1511AB include five main emission lines: Ly$\alpha$, Si\,{\sc 
iv}/O\,{\sc iv}], C\,{\sc iv}, C\,{\sc iii}], and Mg\,{\sc ii}, three of which are also present in 
the LT--SPRAT spectra of the quasar images (see Figures \ref{fig:f3} and \ref{fig:f5}). After 
de--redshifting these spectra to their rest frame (using $z_{\rm{s}}$ = 2.049), we analysed the 
features of interest. The GTC--OSIRIS--R500B data show prominent Ly$\alpha$, Si\,{\sc iv}/O\,{\sc 
iv}], C\,{\sc iv}, and C\,{\sc iii}] emissions in each image, and the corresponding line profiles 
were obtained in a standard way. We took two continuum windows for each emission feature (one on 
its left side and the other on its right), fitted a linear function to the data in both windows, 
and then removed the continuum level in the spectral region. 

In a second step, we performed a multi--component decomposition of these continuum--subtracted line 
profiles. Such a decomposition has been previously used in many spectral studies 
\citep[e.g.,][]{1985ApJ...288...94W,2002ApJS..143..257K,2003ApJ...596..817D,2007A&A...468..885S,
2010MNRAS.409.1033M}. For each emission line, its two profiles (A and B data) were modelled as a 
sum of Gaussian components, i.e., primary components of the line plus other secondary contributions 
blended with primary ones. In a first stage, we only decomposed the A profile by fitting the 
central wavelength of primary components (a single Gaussian or two Gaussians with different widths; 
by setting the wavelength separations between primary and secondary components to known values), as 
well as the widths and amplitudes of all components. For the B profile having lower signal 
strength, the central wavelength of primary components and the widths of all contributions were 
taken from the output of the first stage. Thus, in a second stage, we fitted the B data leaving 
only the amplitudes as free parameters.
 
To reproduce the Ly$\alpha$ profiles, we used a single primary (Ly$\alpha$) component plus a 
red--wing excess due to N\,{\sc v}, while we considered a single component to describe the Si\,{\sc 
iv}/O\,{\sc iv}] profiles (see the top panels of Figure \ref{fig:f8}). The C\,{\sc iv} line 
shapes were well traced by two primary contributions (narrow and broad) plus a He\,{\sc ii} 
complex. This complex is a blend of several lines, and was treated as a secondary Gaussian 
component \citep[e.g.,][]{2002MNRAS.337..275C}. Regarding the C\,{\sc iii}] profiles, we used two 
C\,{\sc iii}] components (narrow and broad), as well as two blue--wing excesses caused by Si\,{\sc 
iii}] and Al\,{\sc iii} \citep[e.g.,][]{1994ApJ...423..131B}. The carbon line profiles and their 
decompositions are shown in the middle panels of Figure \ref{fig:f8}. The GTC--OSIRIS--R500R 
spectra include the Mg\,{\sc ii} emission line, which is not present in the blue grism data. Hence, 
we also carried out the analysis of the Mg\,{\sc ii} emission in each image. We extracted the two 
continuum--subtracted profiles in a standard way (see above), and then did decompositions into a 
single Gaussian component (see the bottom panel of Figure \ref{fig:f8}). To avoid the wings of the 
Mg\,{\sc ii} line that are presumably contaminated by Fe\,{\sc ii}/Balmer emission 
\citep[e.g.,][]{1985ApJ...288...94W}, we exclusively fitted the 40 \AA\ width central region. 

Although we do not detect variations of the Si\,{\sc iv}/O\,{\sc iv}], C\,{\sc iv}, C\,{\sc iii}], 
and Mg\,{\sc ii} emissions in the A image (in Figure \ref{fig:f8}, there is great similarity 
between the GTC--OSIRIS profiles and the SDSS/BOSS line shapes observed three years before), the 
LT--SPRAT spectra allow us to directly compare A and B profiles at the same emission time. These 
spectra were taken at two epochs separated by the time delay between images (see Sections 
\ref{subsubsec:livspec} and \ref{sec:delay}), so we compared the A profiles on 2015 August 16$-$18 
with the B profiles on 2016 March 17. The LT--SPRAT data do not cover the Mg\,{\sc ii} emission, 
and their Si\,{\sc iv}/O\,{\sc iv}] signals were too noisy to be useful. We thus concentrated on 
the carbon line profiles (see Figure \ref{fig:f9}), which were obtained and decomposed as those in 
Figure \ref{fig:f8}. From the right panel of Figure \ref{fig:f9}, we see that the C\,{\sc iii}]
decomposition for the B image is far from robust.

From the profiles and decompositions in Figures \ref{fig:f8} and \ref{fig:f9}, it is 
straightforward to obtain magnitude differences for emission line cores and components. For a given 
emission line (in the GTC--OSIRIS or LT--SPRAT spectra), $(B - A)_{\rm{core}} = - 2.5 \log 
(B/A)_{\rm{core}}$ was estimated by integrating the A and B profiles over a 20 \AA\ width central 
region (line core), while $(B - A)_{\rm{comp}} = - 2.5 \log (B/A)_{\rm{comp}}$ for each of its 
components was derived by integrating the two associated Gaussian distributions. In addition, the 
1$\sigma$ confidence intervals for the magnitude differences were determined from 1000 pairs AB of 
simulated spectra in the region of interest. We obtained a pair AB of simulated spectra in the same 
manner as a pair of synthetic light curves in Section \ref{sec:delay}. Instead of photometric 
errors, here we used the standard deviations of the residual flux in the two continuum windows to 
add random deviations to the observed fluxes. In Table \ref{tab:t7}, we present our measurements of 
$(B - A)_{\rm{core}}$ and $(B - A)_{\rm{comp}}$ having relative errors less than 10\%. We note that
all line components have widths exceeding the instrumental ones of $\sim$ 2$-$3 \AA.

\begin{deluxetable}{lcccccc}
\tablecaption{Analysis of emission lines.\label{tab:t7}}
\tablenum{7}
\tablewidth{0pt}
\tablehead{
\colhead{Instrument} &
\colhead{Main line} &
\colhead{$\lambda_{\rm{rest}}$\tablenotemark{a}} &
\colhead{$(B - A)_{\rm{core}}$\tablenotemark{b}} & 
\colhead{Component} &
\colhead{$\sigma_{\rm{comp}}$\tablenotemark{c}} &
\colhead{$(B - A)_{\rm{comp}}$\tablenotemark{b}} \\  
\colhead{} & \colhead{} & \colhead{(\AA)} & \colhead{(mag)} & 
\colhead{} & \colhead{(\AA)} & \colhead{(mag)} 
}
\startdata
GTC--OSIRIS--R500B & Ly$\alpha$ & 1216.09 $\pm$ 0.15 & 0.590 $\pm$ 0.030 & 
single & 6.37 $\pm$ 0.16 & 0.595 $\pm$ 0.028 \\ 
	& Si\,{\sc iv}+O\,{\sc iv}] & 1399.83 $\pm$ 0.16 & 0.628 $\pm$ 0.037 & 
single & 9.88 $\pm$ 0.17 & 0.614 $\pm$ 0.038 \\ 
	& C\,{\sc iv} & 1549.58 $\pm$ 0.02 & 0.512 $\pm$ 0.003 & 
narrow & 7.25 $\pm$ 0.09 & 0.629 $\pm$ 0.008 \\ 
	&  		  & 			     & 			 & 
broad  & 26.16 $\pm$ 0.33 & 0.398 $\pm$ 0.008 \\ 
	&  		  & 			     & 			 & 
He\,{\sc ii} & 22.89 $\pm$ 0.36 & 0.345 $\pm$ 0.022 \\ 
	& C\,{\sc iii}] & 1909.61 $\pm$ 0.37 & 0.584 $\pm$ 0.013 & 
narrow & 6.15 $\pm$ 0.87 & 0.544 $\pm$ 0.038 \\ 
	&  		  & 			     & 			 & 
broad  & 19.37 $\pm$ 2.72 & 0.608 $\pm$ 0.053 \\ 
GTC--OSIRIS--R500R & Mg\,{\sc ii} & 2800.25 $\pm$ 0.10 & 0.509 $\pm$ 0.010 & 
single & 15.32 $\pm$ 0.15 & 0.507 $\pm$ 0.010 \\ 
LT--SPRAT & C\,{\sc iv} & 1550.43 $\pm$ 0.13 & 0.507 $\pm$ 0.017 & 
narrow & 6.17 $\pm$ 0.34 & 0.681 $\pm$ 0.063 \\ 
& C\,{\sc iii}] & 1910.38 $\pm$ 1.16 & 0.643 $\pm$ 0.057 & 
\nodata & \nodata & \nodata \\  
\enddata
\tablenotetext{a}{Central wavelength of primary components of the line in the rest frame of the 
source ($z_{\rm{s}}$ = 2.049)}
\tablenotetext{b}{Magnitude difference at the same observing time (GTC--OSIRIS) or at the same 
emission time (i.e., magnification ratio; LT--SPRAT). $(B - A)_{\rm{core}}$ is estimated in the 20 
\AA\ wide core (rest frame) of the primary emission, while $(B - A)_{\rm{comp}}$ is associated with 
a primary component (single, narrow or broad) or a secondary one}
\tablenotetext{c}{Rest-frame standard width of the Gaussian component}
\end{deluxetable}

\subsection{Solutions for the visual extinction and macrolens magnification ratios} \label{subsec:dm}

Although the ideal procedure to obtain a reliable extinction--macrolens solution for \object{SDSS 
J1515+1511} is to use pure narrow lines arising from the NLR \citep[e.g.,][]{2003MNRAS.339..607M}, 
there are no available data on this type of emission lines. Thus, in this section, the magnitude 
differences in Table \ref{tab:t7} are used to study the visual extinction and macrolens 
magnification 
ratios in the lens system. The line cores are presumably produced in extended regions that are 
unaffected by microlensing \citep[e.g.,][and references therein]{2012ApJ...755...82M}, and this 
hypothesis is assumed true unless evidence indicates otherwise. We focused on the five lines in the 
GTC--OSIRIS spectra, i.e., magnitude differences at the same observing time, checking through 
previous SDSS/BOSS data of A and the LT--SPRAT results for the carbon lines whether intrinsic 
variability is playing a role. 

For the Ly$\alpha$ and Si\,{\sc iv}/O\,{\sc iv}] lines, their $(B - A)_{\rm{core}}$ values are 
close to and consistent with the magnitude differences for their single primary components (see the 
two first result rows of Table \ref{tab:t7}). We taken the line--core differences as the two first 
data to get extinction--macrolens solutions. For the GTC--OSIRIS C\,{\sc iv} emission line, we have
a richer information, and detect a $B - A$ gradient: $(B - A)_{\rm{narrow}} \sim$ 0.63, 
$(B - A)_{\rm{core}} \sim$ 0.51, and $(B - A)_{\rm{broad}} \sim$ 0.40. In this case, even the line 
core seems to be affected by microlensing, and we considered the $(B - A)_{\rm{narrow}}$ value as 
the third data point for our study of extinction--macrolens parameters. We remark that the 
existence of microlensing in the C\,{\sc iv} line core is also supported by the results derived 
from the LT--SPRAT spectra. For the other carbon line (C\,{\sc iii}]) in the GTC--OSIRIS spectra, 
the $(B - A)_{\rm{narrow}}$ and $(B - A)_{\rm{broad}}$ values have large uncertainties, so they are 
consistent with each other and with the line--core difference. We taken this last difference 
(fourth data point) because it has the smallest error. For the Mg\,{\sc ii} line, we considered its 
$(B - A)_{\rm{core}}$ value (fifth data point), which basically coincides with the difference for 
the single primary component.

\begin{figure}[ht!]
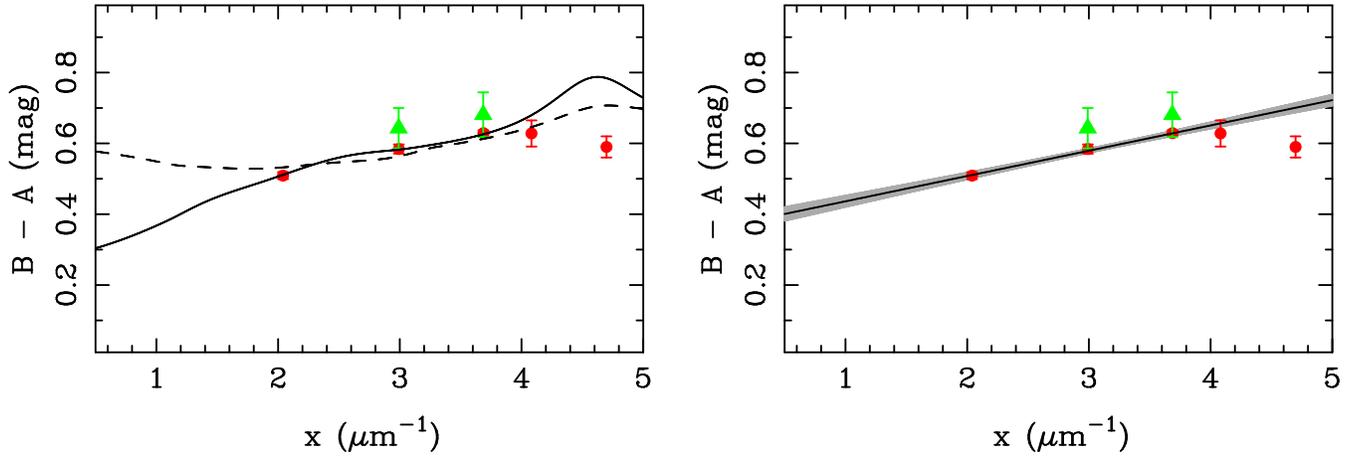

\begin{minipage}[h]{0.5\linewidth}
      \centering
      \includegraphics[angle=-90,width=0.95\textwidth]{MWdust.eps}
      \end{minipage}
	\begin{minipage}[h]{0.5\linewidth}
      \centering
      \includegraphics[angle=-90,width=0.95\textwidth]{STdust.eps}
    	\end{minipage}
\caption{Extinction curve for SDSS J1515+1511. We display seven magnitude differences (see main 
text): five GTC--OSIRIS data for the Ly$\alpha$, Si\,{\sc iv}/O\,{\sc iv}], C\,{\sc iv}, C\,{\sc 
iii}], and Mg\,{\sc ii} emission lines (red circles), and two LT--SPRAT data for the carbon lines 
(green triangles). The x--axis represents the inverse of the wavelength in the rest frame of G1. 
Left: Assuming a Galactic extinction law in G1, the dashed and solid lines describe the best fits 
using all the GTC--OSIRIS data and the first four GTC--OSIRIS differences (excluding the data point 
for the Ly$\alpha$ emission), respectively. Right: Assuming a linear extinction law in G1, we show 
the best fit to the first four GTC--OSIRIS differences (solid line) and its 1$\sigma$ band (light 
grey area).}
\label{fig:f10} 
\end{figure}
 
First, we fitted a Galactic extinction model to the five GTC--OSIRIS magnitude differences that are 
described in the previous paragraph. The model was relied on the general formalism for the 
differential extinction in a pair of lensed images \citep[e.g.,][]{1999ApJ...523..617F,
2003A&A...405..445W,2006ApJS..166..443E}, assuming the presence of Milky Way--like dust 
\citep{1989ApJ...345..245C} in the main lensing galaxy at $z_{\rm{G1}}$ = 0.742. More specifically, 
we used the equation (2) of \citet{2014A&A...568A.116S} to obtain a best fit with $\chi^2 \sim$ 27 
(two degrees of freedom). This poor model fit is largely due to the large residual for the 
Ly$\alpha$ magnitude difference (see the dashed line in the left panel of Figure \ref{fig:f10}). As 
shown in Figure \ref{fig:f8}, despite the similarity between the Si\,{\sc iv}/O\,{\sc iv}], C\,{\sc 
iv}, C\,{\sc iii}], and Mg\,{\sc ii} line profiles from the GTC--OSIRIS and SDSS/BOSS data of the A 
image, the Ly$\alpha$ line strength from the GTC--OSIRIS data of A is substantially smaller than 
that derived through previous SDSS/BOSS observations of the same image (three years before). 
Therefore, there is evidence of Ly$\alpha$ variability in the A image, and this precludes the use 
of the GTC--OSIRIS Ly$\alpha$ magnitude difference in the fits. When fitting the model to 
exclusively the differences at $x = (1 + z_{\rm{G1}})/\lambda \sim$ 2$-$4 $\mu$m$^{-1}$, i.e., 
excluding the Ly$\alpha$ measurement, we found a best fit with $\chi^2 \sim$ 2 (one degree of 
freedom; see the solid line in the left panel of Figure \ref{fig:f10}).  

In spite of the notable improvement in the reduced chi--square value when we do not take the 
Ly$\alpha$ difference into account, the Galactic extinction model does not work satisfactorily 
because $\chi^2_{\rm{r}} \sim$ 2 is still large. Thus, we also fitted a linear extinction model to 
the GTC--OSIRIS Si\,{\sc iv}/O\,{\sc iv}], C\,{\sc iv}, C\,{\sc iii}], and Mg\,{\sc ii} 
differences. The linear extinction law describes reasonably well the dust effects at $x \sim$ 2$-$4 
$\mu$m$^{-1}$ \citep[e.g.,][]{1984A&A...132..389P}, and we used the equation (3) of 
\citet{2014A&A...568A.116S} with $\alpha$ = 1. This standard linear model fits much better than the 
Galactic model, since we obtained $\chi^2 \sim$ 1 with two degrees of freedom. In the right panel 
of Figure \ref{fig:f10}, we present the best fit (solid line) and the 1$\sigma$ band (light grey 
area). As expected, the Ly$\alpha$ difference (last red circle) behaves as an outlier, while there 
is an acceptable agreement between the carbon differences at the same observing time (GTC--OSIRIS)
and the same emission time (LT--SPRAT). The macrolens magnification and visual extinction ratios 
are $\Delta m_{\rm{AB}}$ = 0.365 $\pm$ 0.023 and $\Delta A_{\rm{AB}}(V)$ = 0.130 $\pm$ 0.013 mag, 
respectively (1$\sigma$ intervals). 

It is noteworthy that our 68\% confidence interval for $\Delta m_{\rm{AB}}$ incorporates central 
values of the wider range used by \citet{2016MNRAS.458....2R} to infer their SIE+$\gamma$ lens 
model. To account for time delay, dust extinction, and microlensing effects 
\citep[e.g.,][]{2008A&A...478...95Y}, they considered the $K'$--band magnitude difference with an 
increased error (0.34 $\pm$ 0.11 mag) as a proxy to $\Delta m_{\rm{AB}}$. Furthermore, we are
implicitly assuming that the strong absorber at $z_{\rm{abs}}$ = 0.742 (intervening gas) reported 
by \citet{2014AJ....147..153I} is associated with intervening dust at the same redshift. The new 
GTC spectrum of the B image in the top panel of Figure \ref{fig:f3} also shows this Fe/Mg 
absorption system, which is not detected in the GTC spectrum of the A image. We consistently find 
that the B image is more affected by dust extinction, i.e., $A_{\rm{B}}(V) > A_{\rm{A}}(V)$.      

\subsection{Evidence for quiescent microlensing activity} \label{subsec:micro}

\begin{figure}[ht!]
\centering
\includegraphics[angle=-90,width=0.7\textwidth]{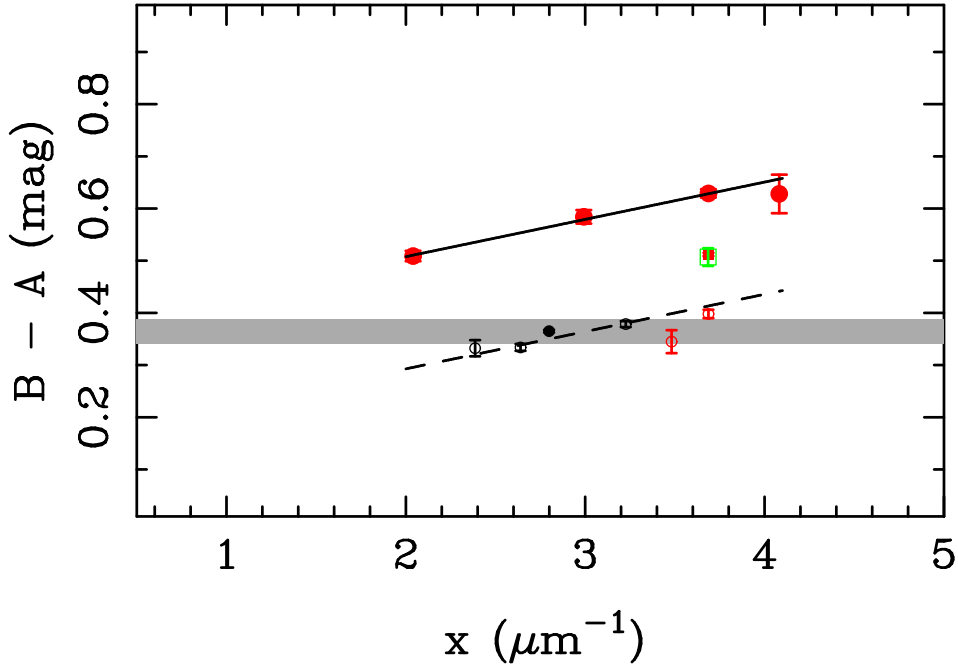}
\caption{Magnification ratios in SDSS J1515+1511. The four filled red circles represent the 
GTC--OSIRIS magnitude differences for the C\,{\sc iv} narrow component, and the Si\,{\sc 
iv}/O\,{\sc iv}], C\,{\sc iii}], and Mg\,{\sc ii} line cores. As reference, the best fit in the 
right panel of Figure \ref{fig:f10} is also drawn (solid line). For the C\,{\sc iv} emission line, 
the filled red and open green squares represent the GTC--OSIRIS and LT--SPRAT line--core 
differences, respectively. We also show the GTC--OSIRIS differences for the C\,{\sc iv} and 
He\,{\sc ii} broad components (open red circles). These two differences and four additional LT 
ratios for the continuum (black circles; see main text) are distributed around a straight line 
(dashed line) parallel to the best extinction--macrolens solution (solid line). The light grey 
rectangle describes the 1$\sigma$ confidence interval for the macrolens magnification ratio.}
\label{fig:f11}
\end{figure}

In Figure \ref{fig:f11}, we show the four GTC--OSIRIS magnitude differences (filled red circles) 
that have been used to obtain the best extinction--macrolens solution (solid line). In addition, 
the light grey rectangle below these circles highlights the 1$\sigma$ confidence interval for the 
macrolens magnification ratio. For the C\,{\sc iv} emission line, we have reported on a $B - A$ 
gradient in Section \ref{subsec:dm}. This supports the presence of microlensing in the C\,{\sc iv} 
BLR, since $B - A$ decreases from $\sim$ 0.63 (GTC--OSIRIS narrow component that is originated in a 
very extended region; third filled red circle in Figure \ref{fig:f11}) to $\sim$ 0.40 (GTC--OSIRIS 
broad component arising from a relatively compact BLR; open red circle under the third filled red 
circle), passing through an intermediate value of $\sim$ 0.51 (GTC--OSIRIS and LT--SPRAT line--core 
differences; filled red and open green squares). Regarding the He\,{\sc ii} (broad) component in 
the GTC--OSIRIS spectra, its $(B - A)$ value is slightly less than $(B - A)_{\rm{broad}}$ for the 
C\,{\sc iv} emission (see the two open red circles), which is an additional evidence for 
microlensing in the high--ionisation BLR \citep[e.g.,][and references 
therein]{2013ApJ...764..160G}. 

In Figure \ref{fig:f11}, we also display LT magnification ratios for the continuum at different 
wavelengths: the filled black circle represents our 1$\sigma$ measurement of $\Delta r_{\rm{AB}}$ 
in Section \ref{sec:delay}, while the three open black circles represent 1$\sigma$ measurements 
from LT--SPRAT data in three 200 \AA\ wide spectral intervals centred at 5400, 6600, and 7300 \AA. 
These black circles suggest that the continuum emitting region (accretion disk) and the 
high--ionisation BLR suffer a similar microlensing effect of $\sim$ 0.2 mag (the dashed line serves 
as a guide to the eye). Our observations in both the wavelength and time domains indicate that the 
compact sources are crossing microlensing magnification regions without appreciable gradients.  
 
\section{Lensing mass distribution} \label{sec:lensmod}

\citet{2016MNRAS.458....2R} used the relative astrometry of SDSS J1515+1511ABG1 and the magnitude 
difference $B - A$ (see Table 3 of that paper), as well as the observed ellipticity and orientation 
of G1 (see Table 4 of that paper), to obtain a SIE+$\gamma$ lens model. These observational 
constraints were inferred from high--resolution $K'$--band imaging on 2012 February 21, and the 
mass of G1 was reasonably modelled as a SIE that is aligned with the observed light distribution. 
Within such a framework, the SIE has only one free parameter (Einstein radius $\theta_{\rm{Ein}}$), 
and the galaxies outside the strong lensing region produce an external shear that is characterised 
by two additional free parameters: shear strength ($\gamma$) and direction ($\theta_{\gamma}$). 
Because the number of model parameters was the same as the number of observational constraints, 
Rusu et al. obtained a perfect fit with $\chi^2 \sim$ 0.

While the Rusu et al.'s results for the lensing mass parameters were derived through the GLAFIC 
software \citep{2010PASJ...62.1017O}, we used the LENSMODEL software \citep{2001astro.ph..2340K} to 
analyse the lens system. Although different software packages may lead to different output results 
for the same lens scenario \citep[e.g.,][]{2015arXiv150500502L}, we reproduced the Rusu et al.'s 
GLAFIC best--solution for the SIE+$\gamma$ scenario by using the LENSMODEL package (see the second 
column of Table \ref{tab:t8}). The GLAFIC and LENSMODEL definitions of $\theta_{\rm{Ein}}$ for a 
SIE differ by a factor $f(q) = [(1 + q^2)/2q]^{1/2}$, where $q = b/a = 1 - e$ is the axis ratio and 
$e$ is the ellipticity. Thus, in Table \ref{tab:t8}, we quote the values of 
$\theta_{\rm{Ein}}{\rm(GLAFIC)} = \theta_{\rm{Ein}}{\rm(LENSMODEL)} \times f(q)$. Instead of the 
Rusu et al.'s constraint on $\Delta m_{\rm{AB}}$ (or equivalently, on the macrolens flux ratio 
$B/A$), we considered the new accurate measurement in Section \ref{subsec:dm}. Moreover, we also 
incorporated the measured time delay (Section \ref{sec:delay}) as an additional constraint. This 
allowed us to fit the redshift of the lensing mass $z_{\rm{l}}$ by assuming a concordance cosmology 
with $H_0$ = 70 km s$^{-1}$ Mpc$^{-1}$, $\Omega_m$ = 0.27, and $\Omega_{\Lambda}$ = 0.73 
\citep{2009ApJS..180..330K}. Regarding the source redshift, we taken $z_{\rm{s}}$ = 2.049 rather 
than 2.054 (see discussion in Section \ref{sec:intro}). 

\begin{deluxetable}{ccc}
\tablecaption{SIE+$\gamma$ mass model.\label{tab:t8}}
\tablenum{8}
\tablewidth{0pt}
\tablehead{
\colhead{Parameter} & \colhead{Rusu et al. (2016)} & \colhead{This paper}  
}
\startdata 
$\theta_{\rm{Ein}}$        & $1\farcs 21$         & $1\farcs 21$         \\           
$e$                   	   & (0.81)               & (0.81)               \\
$\theta_e$          	   & ($-17\fdg 1$)        & ($-17\fdg 1$)        \\
$\gamma$                   & 0.283                & 0.286                \\
$\theta_{\gamma}$          & $76\fdg 0$           & $76\fdg 0$           \\
$z_{\rm{l}}$               & \nodata              & 0.729                \\
$\chi^2$     	         & $\sim$ 0             & $\sim$ 0             \\ 
\enddata
\tablecomments{Here, $\theta_{\rm{Ein}}$, $e$, $\gamma$, and $z_l$ denote Einstein radius, 
ellipticity, external shear strength, and lens redshift, respectively; and position angles 
($\theta_e$ and $\theta_{\gamma}$) are measured east of north. The quantities within parentheses 
were not fitted, but fixed at values derived from the light distribution of G1.}
\end{deluxetable}

Our best values of $\theta_{\rm{Ein}}$, $\gamma$, and $\theta_{\gamma}$ (see the third column of 
Table \ref{tab:t8}) just about match those of the Rusu et al. Concerning the lens redshift, its 
best value is slightly lower than 0.742, which has been used to obtain the constraint on the 
macrolens magnification ratio in Section \ref{subsec:dm}. However, the formal 1$\sigma$ interval 
(0.729 $\pm$ 0.011 for $\Delta \chi^2 \leq 1$) is consistent with $z_{\rm{l}} \sim$ 0.74, and 
the best value of $z_{\rm{l}}$ should be increased in presence of an external convergence (see 
discussion at the end of this section). Additionally, a moderately high value of $H_0$ = 72 km 
s$^{-1}$ Mpc$^{-1}$ \citep[studies of Cepheids, maser galaxies, supernovae, gravitational lenses, 
and other astrophysical objects support $H_0 \sim 72-74$ km s$^{-1}$ Mpc$^{-1}$; 
e.g.,][]{2010ARA&A..48..673F,2015LRR....18....2J} led to $z_{\rm{l}}$ = 0.742. Despite our failure 
in accurately solving the G1 spectrum and to measure its redshift directly, it is easy to reconcile 
the lens redshift with the assumed redshift of G1. This last result strongly supports that G1 is 
located at $z_{\rm{G1}}$ = 0.742.  

We remark that dark matter halos of some galaxies other than G1 could extend to the strong lensing 
region, so the SIE+$\gamma$ scenario could be not so realistic as it seems. Apart from the absorber 
most likely associated with G1, the top panel of Figure \ref{fig:f3} displays a more distant Fe/Mg 
absorption system ($z_{\rm{abs}}$ = 1.166) that is seen in both quasar images, but affecting A in a 
more significant manner. The secondary galaxy G2 may also play a role. For instance, taking into 
account $z_{\rm{G2}}$ = 0.541, as well as the angular separation between this galaxy and the quasar 
images $\theta \sim 15\arcsec$, we found that the dark matter halo of G2 may reach the region of 
interest at $\sim$ 100 kpc. Therefore, using a singular isothermal sphere (SIS) to describe 
the mass of G2, we studied the possible gravitational effect at the position of the double quasar. 
Assuming that $K$(G2) $\sim$ $K'$(G2) = 17.4 mag \citep{2016MNRAS.458....2R}, the lensing 
Faber--Jackson relation \citep{2003ApJ...587..143R} yielded a dark matter velocity dispersion 
$\sigma_{\rm{DM}}$(G2) $\sim$ 196 km s$^{-1}$. If $z_{\rm{l}}$ were equal to $z_{\rm{G2}}$, the SIS 
would produce a convergence and shear of $\kappa = \gamma \sim$ 0.023 at $\theta \sim 15\arcsec$. 
However, as G2 lies at a redshift different to $z_{\rm{l}}$, the effective convergence and shear 
would be $\kappa_{\rm{eff}} = \gamma_{\rm{eff}} \sim$ 0.015 \citep{2006ApJ...641..169M}. Thus, G2 
may account for only a small fraction of the external shear in Table \ref{tab:t8}, while it could 
be responsible of a slight increase in $z_{\rm{l}}$ \citep[see 
discussion in][]{2010ApJ...708..995G}.   

\section{Summary and conclusions} \label{sec:end}  

This analysis uses new LT light curves (and spectra) and GTC spectra of the gravitational lens 
system \object{SDSS J1515+1511} to measure the time delay between its two quasar images (A and B), 
as well as to discuss effects and physical properties of intervening objects. All optical data 
correspond to observations over the last 3 years. Our main results and conclusions are (error bars 
represent 1$\sigma$ confidence intervals):

1) We find that the lensed quasar lacks microlensing activity. This means that the accretion disk 
and the inner BLR are suffering an almost constant differential magnification by microlenses 
(stars) in the main lensing galaxy G1. 

2) The intrinsic fluctuations seen in the $r$--band LT light curves of A and B (with an amplitude 
of $\sim$ 0.1--0.3 mag) lead to a robust time delay $\Delta t_{\rm{AB}}$ = 211 $\pm$ 5 days (A is 
leading). In the current quiescent state of microlensing activity, \object{SDSS J1515+1511} is 
particularly well suited for reverberation mapping studies. After detecting a prominent event in an 
optical light curve of A, one has several months to prepare a multiwavelength monitoring of B 
\citep[e.g.,][]{2012ApJ...744...47G,2015ApJ...813...67D}. 

3) Our GTC data do not allow us to extract an accurate spectrum of G1, which is fainter than 22 mag 
in the $r$ band. However, Fe/Mg absorption features in the GTC--OSIRIS--R500B spectrum of the 
quasar image closer to G1 (B image) suggest that $z_{\rm{G1}}$ = 0.742 \citep[see the top panel of 
Figure \ref{fig:f3} and earlier findings by][]{2014AJ....147..153I}. Assuming this redshift for G1, 
we carefully analyse the differential extinction in G1 and the macrolens magnification ratio 
$\Delta m_{\rm{AB}}$. From the main emission lines in the GTC spectra of the quasar (with the help 
of LT and SDSS/BOSS spectra), we infer a visual extinction ratio of 0.130 $\pm$ 0.013 mag (B is 
redder), in agreement with the presence of more dust where there is more gas. In addition, we 
obtain $\Delta m_{\rm{AB}}$ = 0.365 $\pm$ 0.023 mag. 

4) We use previous observational constraints on the relative astrometry of the lens system and the 
morphology of G1 \citep{2016MNRAS.458....2R}, together with the new constraints on the time delay 
and the macrolens magnification ratio, to update the Rusu et al.'s SIE+$\gamma$ lens model. Our 
results for the mass scale of the SIE (G1) and the external shear practically coincide with those 
of Rusu et al., and using a standard concordance cosmology, we derive an 1$\sigma$ interval for the 
lens redshift $z_{\rm{l}}$ that is marginally consistent with the assumed redshift of G1. 
Additionally, there is some evidence for the existence of a small external convergence (see 
below) leading to a better agreement between $z_{\rm{l}}$ and $z_{\rm{G1}}$. We also obtain 
$z_{\rm{l}}$ = 0.742 for $H_0$ = 72 km s$^{-1}$ Mpc$^{-1}$ \citep[e.g.,][]{2010ARA&A..48..673F,
2015LRR....18....2J}. Thus, our mass modelling confirms the tentative value of $z_{\rm{G1}}$ in the 
discovery paper \citep{2014AJ....147..153I}. 

5) It should be noted that a SIE+$\gamma$ lens scenario is not the only possible. Both quasar 
images intercept a Fe/Mg absorption system at a redshift of 1.166, which may play a role. We also 
measure the redshift of the secondary galaxy G2, which is nearer than G1 and has a dark matter 
velocity dispersion of about 200 km s$^{-1}$ \citep[SIS model; using the scheme 
of][]{2003ApJ...587..143R}. If the mass distribution in G2 would extend up to $\sim$ 100 kpc, then 
it would produce $\sim$ 5\% of the external shear and an external convergence of $\sim$ 0.015. 
 
\acknowledgments

We thank the anonymous referee for her/his helpful comments and suggestions. The Liverpool 
Telescope is operated on the island of La Palma by Liverpool John Moores University 
in the Spanish Observatorio del Roque de los Muchachos of the Instituto de Astrofisica de Canarias 
with financial support from the UK Science and Technology Facilities Council. This article is also 
based on observations made with the Gran Telescopio Canarias (GTC), installed at the Spanish 
Observatorio del Roque de los Muchachos of the Instituto de Astrof\'{\i}sica de Canarias, in the 
island of La Palma. We thank the staff of both telescopes for a kind interaction before, during and 
after the observations. We also used data taken from the SDSS databases, and we are grateful to the 
SDSS collaboration for doing those public databases. This research has been supported by the 
Spanish Department of Research, Development and Innovation grant AYA2013-47744-C3-2-P 
(Gravitational LENses and DArk MAtter - GLENDAMA project), and the University of Cantabria.



\vspace{5mm}
\facilities{Liverpool:2m(IO:O and SPRAT), GTC(OSIRIS)}

\software{IRAF, IMFITFITS, LENSMODEL}

\end{document}